\documentclass[preprint,12pt]{elsarticle}

\usepackage{amssymb}
\usepackage{amsmath,stackrel}
\usepackage{url}
\usepackage[dvipsnames]{xcolor}

\journal{}

\begin{document}

\begin{frontmatter}

\title{Accurate analysis of the pitch pulse-based magnitude/phase structure of natural vowels and assessment of three lightweight time/frequency voicing restoration methods\tnoteref{thanks1}}
 \tnotetext[thanks1]{This work was financed by FEDER - Fundo Europeu de Desenvolvimento Regional funds through the COMPETE 2020-Operacional Programme for Competitiveness and Internationalization (POCI), and by Portuguese funds through FCT-Funda\c{c}\~{a}o para a Ci\^{e}ncia e a Tecnologia in the framework of the project POCI-01-0145-FEDER-029308.}

\author[1]{An\'{\i}bal J. S. Ferreira\corref{cor1}}
\ead{ajf@fe.up.pt}
\author[2]{Luis M. T. Jesus}
\ead{lmtj@ua.pt}
\author[3]{Laurentino M. M. Leal}
\ead{desmen47@hotmail.com}
\author[3]{Jorge E. F. Spratley}
\ead{jorge.spratley@gmail.com}

\affiliation[1]{organization={INESC TEC; Department of Electrical and Computer Engineering, University of Porto -- Faculty of Engineering},
            addressline={Rua Dr. Roberto Frias, s/n}, 
            city={Porto},
            postcode={4200-465}, 
            country={Portugal}}
 \affiliation[2]{organization={Intelligent Systems Associate Laboratory (LASI), Institute of Electronics and Informatics Engineering of Aveiro (IEETA) and School of Health Sciences (ESSUA)},
            addressline={University of Aveiro, Campus Universit\'{a}rio de Santiago}, 
            city={Aveiro},
            postcode={3810-193}, 
            country={Portugal}}
\affiliation[3]{organization={RISE-Health; Department of Otorhinolaryngology, Centro Hospitalar Universit\'{a}rio de S\~{a}o Jo\~{a}o; Department of Surgery and Physiology, University of Porto -- Faculty of Medicine},
            addressline={Alameda Prof. Hern\^{a}ni Monteiro}, 
            city={Porto},
            postcode={4200-319}, 
            country={Portugal}}

\cortext[cor1]{Corresponding author}

\begin{abstract}
Whispered speech is produced when the vocal folds are not used, either intentionally, or due to a temporary or permanent voice condition.
The essential difference between natural speech and whispered speech is that periodic signal components that exist in certain regions of the former, called voiced regions, as a consequence of the vibration of the vocal folds, are missing in the latter.

The restoration of natural speech from whispered speech requires delicate signal processing procedures that are especially useful if they can be
implemented on low-resourced portable devices, in real-time, and on-the-fly, taking advantage of the established source-filter paradigm of voice production and related models. 
This paper addresses two challenges that are intertwined and are key in informing and making viable this envisioned technological realization.

The first challenge involves characterizing and modeling the evolution of the harmonic phase/magnitude structure of a sequence of individual pitch periods in a voiced region of natural speech comprising sustained or co-articulated vowels. This paper proposes a novel algorithm segmenting individual pitch pulses, which is then used to obtain illustrative results highlighting important differences between sustained and co-articulated vowels, and suggesting practical synthetic voicing approaches.

The second challenge involves model-based synthetic voicing. Three implementation alternatives are described that differ in their signal reconstruction approaches: frequency-domain, combined frequency and time-domain, and physiologically-inspired separate filtering of glottal excitation pulses individually generated. The three alternatives are compared objectively using illustrative examples, and subjectively using the results of listening tests involving synthetic voicing of sustained and co-articulated vowels in word context. 
\end{abstract}


\end{frontmatter}


\section{Introduction}
\label{sec:intro}

\subsection{Context, motivation and paper structure}
\label{sec:context}

Voice/speech communication is the most important modality of human social and professional interaction \cite{speech:fant70, speech:shaw2000}. Thus,
enjoying healthy voice production, and natural voice/speech communication, is a condition for self-esteem, well-being, social inclusion, and personal realization. In particular, healthy voice/speech production depends critically on the good physiological condition and operation of the vocal folds in the larynx as their vibration -a process known as phonation- is of paramount importance allowing a voice signal to enhance the phonetic information, project acoustically, and convey an idiosyncratic sound signature.

On the contrary, if the physiological conditions are such that the vibration pattern of the vocal folds is too faint -a condition known as asthenia-,
or if the vocal folds do not vibrate at all -a condition known as aphonia-, then the voice signal sounds muffled and is easily corrupted by competing acoustic signals. As a consequence, the voice signal is not able to project acoustically, the underlying phonetic information does not stand out, nor does a clear sound individuality. It is estimated that the prevalence of voice disorders in the general population, including asthenia and aphonia, is between 1\% and 15\%, affecting more significantly women and teachers \cite{voice:lyberg2005}.

The voice signals that support speech communication consist of time-varying time-frequency representations combining two main categories of signal components.
One category exhibits a periodic structure as a consequence of phonation. This category of signal components is called ``voiced'' and includes sustained
and co-articulated vowel sounds, as well as certain consonants.
The second category of signal components is characterized by an impulsive structure, or a random structure, and results from either a sudden release of air, or from turbulent air, coming from the lungs and passing through a constriction in the larynx, or downstream in the vocal tract, e.g., the teeth.
This category of non-periodic signal components is called ``unvoiced'' and includes specific types of consonants, e.g., plosives and fricatives \cite{speech:rabiner93}.

In natural speech, voiced sounds typically predominate over unvoiced sounds. The periodic character of voiced sounds is beneficial in several regards. First, given that periodicity in the time domain implies a harmonic structure in the spectral domain \cite{sgnps:opp96}, room is created for a robust spectral diversity that significantly contributes to voice projection and discrimination. Second, that harmonic spectral structure plays a very important role in shaping and enhancing vocal tract resonances, which emphasizes the phonetic information, and improves speech intelligibility. Third, idiosyncrasies of a speaker voice sound signature stand out due to peculiar traits of his/her vocal fold operation \cite{speech:ajf_isspit2014}, and peculiar and persistent characteristics of his/her vocal tract resonances \cite{forensics:dellwo2007}.

In specific cases of voice pathologies, including cases of moderate/partial laryngectomy or functional/psychological disturbances, such as spasmodic dysphonia,
voice patients are prevented from phonating in a normal way, or at all \cite{voice:doyle2005}. In this case, which is central to our research context, a type of speech is produced that is known as whispered speech. Here, we assume that with exception for the vocal folds, the articulators of voice patients are functional and that the ability of breathing through the oral and nasal cavities is preserved.

\begin{figure}[htb]
\centering
\includegraphics[width=0.85\columnwidth]{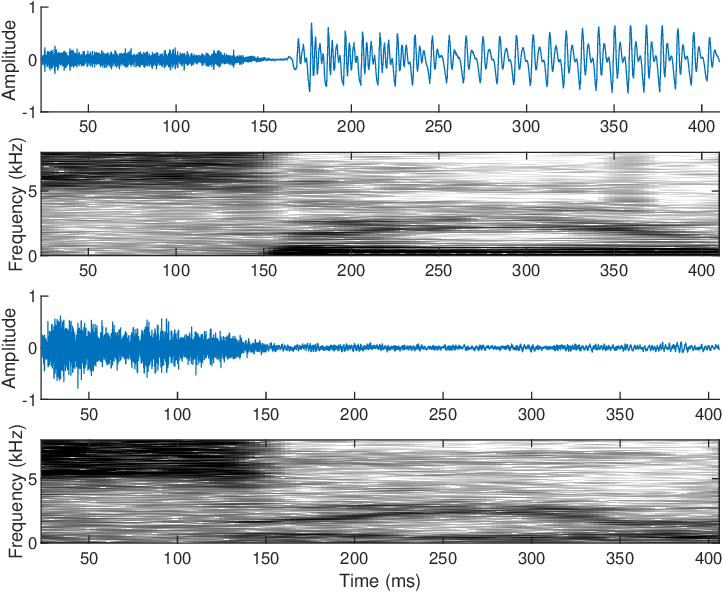}
\caption{From top to bottom: time wave and spectrogram of a voiced speech realization of the Portuguese word `Saiu' by a male speaker; time wave and spectrogram of a whispered speech realization of the same word by the same speaker. In the spectrograms, darker colors denote higher Power Spectral Density levels. The sampling frequency is 22050 Hz.}
\label{fig:SPM06_43_01_saiu}
\end{figure}

Healthy speakers sometimes also use whispered speech deliberately to communicate when privacy is desired or discretion
is recommended. In whispered speech, all phones -the physical materialization of phonemes- are unvoiced, i.e., all acoustic realizations of phonemes consist of unvoiced impulsive and noise-like signal components. Figure \ref{fig:SPM06_43_01_saiu} shows the time waves and associated spectrograms of two voice realizations of the Portuguese word `Saiu' by a male speaker, one consisting of a voiced speech realization (first two plots), another consisting of a whispered speech realization (last two plots).
As Fig. \ref{fig:SPM06_43_01_saiu} shows that the illustrated whispered speech realization does not contain any signal periodicity. Still, the whispered speech contains a number of cues making it intelligible. For example, as the spectrograms in Fig. \ref{fig:SPM06_43_01_saiu} denote, in both voiced and whispered speech realizations the consonant `s' has a strong Power Spectral Density for frequencies above 5 kHz, and the sequence of co-articulated vowels `aiu' is characterized by comparable formant frequencies trajectories.

In practice, however, whispered speech is viable only for very short-distance oral communication given that it is easily corrupted by competing signals or noises, as noted earlier.
Moreover, as common daily life situations involve challenging and strongly time-varying acoustic conditions,
simple voice amplification is not a good solution in an attempt to project whispered speech or to make it more intelligible. Instead, only the regions of whispered speech that correspond to voiced regions in natural speech should be repaired by synthesizing and restoring the missing periodic signal components, as illustrated in Fig. \ref{fig:SPM06_43_01_saiu}. 

Several studies indicate that Portuguese unvoiced phones in natural speech do not differ appreciably from the corresponding phones in whispered speech, if produced by the same speaker \cite{jesus2023, voice:silva2021}. Similar conclusions have been reached in other studies \cite{voice:itakura2005}. Thus, conceptually, we regard unvoiced speech as a baseline signal upon which we may ``implant'' carefully synthesized periodic signal components so as to restore natural speech. In other words, we combine synthetic voicing with the whispered speech so as to restore the sound of voiced/natural speech. In this process,
the periodic components must be correctly shaped in time and frequency so as
to reinforce the intended phonemic/linguistic meaning, to convey a desired individual voice sound signature, and to project acoustically the voice signal.  

This approach has two key advantages. On the one hand, by preserving in the restored speech signal significant portions of the original whispered speech, a relevant degree
of the individual voice sound signature of a given speaker is preserved as it depends on the individual manner unvoiced phonemes are produced, notably plosives and unvoiced fricatives, e.g., sibilants. On the other hand, this approach facilitates combining real-time processing with on-the-fly operation, as explained next.

The broad research and development context motivating the research reported in this paper is the design of a lightweight assistive technology
that is able to restore natural speech from whispered speech, not only in real-time, but also on-the-fly. By real-time we mean that the signal processing underlying the voice restoration process should be faster than the duration of the voice signal being restored. By on-the-fly we mean that relative to the instant a whispered speech sound is produced, the restored voice signal should be presented acoustically within a delay that is commensurate with the maximum delay preventing the perceptual phenomenon of acoustic echo from occurring (i.e., a replica of a sound being perceived as a separate sound), and preventing the problematic lip-sync perceptual phenomenon from occurring (i.e., the image of the lips producing a voice sound being noticeably time-shifted relative to the instant it is heard, which generates perceptual confusion). Typically, the upper limit for that maximum delay is in the order of 50-100 milliseconds \cite{psy:zwi90, psy:moo89}. This represents the maximum allowed delay in the complete processing chain, which includes signal acquisition, analysis, processing, synthesis, and reconstruction, as well as the necessary buffering stages.

A classical technology still being used today mainly by voice patients that went through laryngectomy, or that for some health condition lost the ability to phonate, is the electrolarynx (EL) \cite{voice:cox2019}. It consists of a simple hand-held electronic device that generates a vibratory sound pattern, a kind of sound source that replaces the vibration of the (missing) vocal folds. When the device is pressed against the throat, and the voice patient uses the articulators in a normal way, an artificial periodic -but buzzy- sound is produced that conveys a significant degree of the spectral formant information due to the vocal tract resonances. This is also true if the EL is used with whispered speech. However, the resulting speech sounds unnatural, robotic, and unpleasant because the vibration pattern is monotonous and affects all phonemes. More recent EL models allow the user to manually control the activation of the artificial voicing, and the variation of the fundamental frequency of the artificial tone \cite{voice:cox2019, voice:cohan2022}.

Other than the EL, to our best knowledge, research and development results addressing whispered speech to natural voice conversion have been presented in the literature that comply with the real-time requirement, but not with the on-the-fly requirement. This is discussed further in Subsection \ref{sec:whisper2natspeech}.

Our purpose in this paper is not to describe a complete signal processing system allowing the restoration of natural voice from whispered speech, but rather to
present new results on two key (connected) aspects determining the quality of synthetic voicing: i) the phase and magnitude structure of sustained and co-articulated vowels on a pitch pulse-by-pitch pulse basis, and ii) the objective and perceptual quality of three alternative synthetic voicing strategies for sustained and co-articulated vowels. These new results follow our previous research in the area \cite{voice:silva2021, jesus2023, speech:AJF2024, speech:silvaAJF2020, speech:ajf2020_1, speech:silva2020}.

The remainder of this paper is structured as follows. In Section \ref{sec:pitchpulsesegm} we address previous work that is representative of research on
pitch pulse segmentation in voiced regions of speech. In Section \ref{sec:whisper2natspeech} we provide a high-level overview of representative whispered speech
to natural voice conversion techniques, highlighting those that, similarly to ours, are inspired by the source-filter model-based paradigm.
Section \ref{sec:pitchpulse} describes a new phase-based approach to pitch pulse-by-pitch pulse segmentation in voiced regions of sustained and
co-articulated vowels, and presents illustrative results.
Section \ref{sec:experiments} details three synthetic voicing alternatives that differ in the time-frequency approach to synthesizing periodic signal components 
complying with a desired fundamental frequency contour and magnitude and phase spectral evolution. Objective results are presented that help characterize the
quality of the synthetic signals resulting from each signal processing alternative, as well as results of listening tests that help to characterize their perceptual impact.
Finally, Section \ref{sec:conclusion} summarizes the main results presented and discussed in this paper and addresses future developments.

\subsection{Overview of pitch pulse segmentation}
\label{sec:pitchpulsesegm}

In the literature, the segmentation of individual pitch pulses in a voiced region of speech is required to assist signal processing operations such as high-resolution pitch tracking, the estimation of voice perturbation parameters such as jitter and shimmer, pitch-synchronous speech enhancement and time-scale modification of speech, glottal source waveform estimation, speech prosody analysis, voice conversion, and speech synthesis \cite{voice:cabral2011, voice:harris1993, voice:kubin2006, voice:dikshit2005, voice:cheng1989, voice:drugman2012}.
In many cases, the segmentation process is implemented by locating epoch instants in speech, i.e., those instants associated with sudden alterations of the glottal flow volume velocity, and corresponding to the glottal closure or opening \cite{voice:anan1979, voice:cabral2011}. As the glottal excitation results from the derivative of the glottal flow volume velocity \cite{speech:fant70}, epoch instants are manifested as impulses in the glottal excitation signal that are very important perceptually and are regarded as ``the most significant excitation moments due to the vibration of the vocal folds'' \cite{voice:cabral2011}.

Approaches to the segmentation of individual pitch pulses can be as simple as using the normalized cross-correlation directly applied to a voiced region
of the speech \cite{voice:boersma2005}, to more elaborate procedures involving Linear Prediction (LP) analysis and subsequent normalized cross-correlation
applied to the LP residual, possibly with additional levels of sophistication by including, for example, wavelet analysis, analysis of group delay functions,
Hilbert envelope analysis and linear programming \cite{voice:cabral2011, voice:anan1979}, maximum-likelihood epoch determination \cite{voice:cheng1989},
or Poincar\'e plane analysis \cite{voice:kubin2006}. In several cases, the joint analysis of the speech and electroglottographic (EGG) signals is considered for additional
robustness \cite{voice:cabral2011, voice:legat2011}. Informative reviews of representative pitch pulse segmentation algorithms and approaches can be found in \cite{voice:cabral2011, voice:drugman2012}. 

Some of the practical difficulties that are associated with the reviewed algorithms are as follows.
\begin{itemize}
\item While it is consensual that epochs consist of the most important instants of the glottal excitation and can be taken as reference marks of the start of individual pitch pulses, their practical detection varies significantly according to different authors. For example, the source signal used to estimate and detect them can be the speech signal itself, the residual after inverse LP filtering of the speech signal, or the EGG signal. As the polarity of each of these signals may change with time, especially in regions of co-articulated vowels, uncertainty is added to the estimation process that may give rise to estimation errors. Moreover, as pointed out by Ananthapadmanabha and Yegnanarayana, ``the all-pole model implicitly assumes a minimum phase characteristic for the speech signal, if this is not valid, the phase response of the vocal tract system is not compensated exactly by the digital inverse filter'' \cite{voice:anan1979}, which creates an additional type of uncertainty in the signal analysis and estimation process.
\item Several authors note that it is not unfrequent for a single pitch pulse to contain more than one epoch, which may arise due to the impact of glottis closure or glottis opening \cite{voice:cabral2011, voice:anan1979}. This, again, creates uncertainty in the signal analysis and epoch estimation.
\item Other authors have noted that in the analysis of EGG signals, in some cases, extra glottal pulses appear in a single pitch period making that more than two epochs show up in a single period of the voice signal \cite{voice:henrich2005}. This calls for new and more robust approaches segmenting pitch pulses that reflect the true periodicity in a voiced region of a speech or singing signal.
\end{itemize}
The above arguments warrant an entirely new approach to pitch pulse estimation that can overcome the indicated pitfalls. However, in this paper, more that finding the detailed pitch pulse variations, we want to delimit exactly each pitch pulse in order to study the evolution of the spectral magnitude and phase structure of successive pitch pulses, not only in the case of sustained vowels, but more importantly in the case of co-articulated vowels. To our best knowledge, no such research results have been previously published in the literature. In Section \ref{sec:pitchpulse}, we describe an approach complying with these objectives and that is based on the segmentation of pitch periods -{\em in lieu\/} of epochs- by analyzing the phase of the fundamental frequency.

\subsection{Overview of natural speech reconstruction from whispered speech}
\label{sec:whisper2natspeech}

Two main classes of technical approaches implementing reconstruction of natural/voiced speech from whispered speech can be devised: those that follow a model-based approach, and those that follow a data-driven approach.

\subsubsection{Model-based approaches}
\label{sec:whisper2natspeech_MB}

Representative model-based approaches addressing whispered speech to natural speech conversion and that explicitly exploit domain
knowledge according to the classic source-filter theory of voice production \cite{speech:fant70} are reviewed in \cite{voice:perrotin2020, voice:ajf2016}.

Pioneering work in model-based whispered speech to voiced-speech conversion was carried out by Morris and Clements \cite{speech:morris2002}. Their approach was based on a Mixed-Excited Linear Prediction (MELP) speech coding algorithm in which the LPC residual is a synthetic signal whose fundamental frequency ($f_{0}$) is controlled by specific formant frequencies that are estimated from the whispered speech. The formant frequencies are shifted in frequency accounting for differences between the spectral envelopes of voiced speech and whispered speech with the same phonetic content.
Other authors also developed similar approaches using alternative speech codecs (e.g., the Code Excited Linear Prediction -CELP- codec) \cite{speech:ian2015, speech:ian2013}. These approaches are, however, highly constrained in the signal reconstruction stage due to the codec structure. In fact, in general, the excitation signal is of poor quality because its synthetic version fails to be fully consistent with the analysis/synthesis framework of the codec. This, in turn, impacts negatively on the quality and diversity of the generated speech sounds. In addition, the speech processing rules tend to be the same for different phonemes which further sacrifices phoneme specificity and sound quality. Overall, the subjective quality of the synthetic speech is poor, unpleasant, and speaker-specific features are not preserved or convincingly restored in the synthetic voiced speech. Slightly better results have been reported in \cite{speech:hamid2010}, where several improvements were introduced, including a phoneme classification rule that excludes from voicing regions of the whispered speech that are classified as silence, plosives or unvoiced fricatives.
This is a very important design aspect that we also adopted in \cite{voice:ajf2016}.
A variation of the model-based approaches involves the use of statistical-based tools to map whispered-speech features into voiced-speech features prior to speech reconstruction \cite{voice:alku2018}. A few studies use Gaussian Mixture Modeling (GMM) but tend to suffer from low speech quality and unnatural prosody 
\cite{speech:toda2012}.
In general, statistical-based methods rely on massive training and on a very good match between `source' and `target' data. Typically, they lack flexibility in the synthesis process, and the underlying system complexity is neither amenable for implementation on portable devices, nor suited for real-time operation.

More recently, Perrotin and McLoughlin have proposed a model-based method for whispered speech to natural speech conversion \cite{voice:perrotin2020}.
The proposed algorithm replaces the noisy whisper sound source, in selected regions of the speech, with a synthesized speech-like harmonic source, while keeping the vocal tract spectral envelope unchanged. The replacement is controlled by a parameter assessing the center of gravity of the spectrum of the whispered speech so as to prevent unvoiced phonemic regions from undergoing artificial voicing. The input whispered speech signal is first decomposed into two filters, a vocal tract filter and a glottis filter, in addition to the glottal excitation signal.
The glottis filter is modeled according to a 3rd-order all-pole model: two complex-conjugate poles model the glottal formant, and one pole models the glottal spectral tilt. This model is enforced  in both unvoiced and voiced regions of the speech signal, the difference lying in the glottis filter coefficients that are adapted differently according to those regions.
Synthetic voicing is achieved by generating first a spectrally flat pulse train through additive synthesis where random magnitude and phase perturbations are introduced in order give the synthetic sound a certain degree of naturalness.

Structurally, the Perrotin and McLoughlin model-based proposal possesses several critical limitations, as described next.
\begin{itemize}
\item First, the signal decomposition and reconstruction approach is not perfect reconstructing, which means that those regions of the whispered speech input signal that should not be modified, namely plosives and unvoiced consonants, are likely to suffer signal degradation.
\item Second, voicing depends on a rigid model of a spectrally flat pulse structure, and the glottis filter is constrained to a 3rd-order all-pole model that is the same in both voiced and unvoiced regions of the synthetic signal. This constrains the ability to generate flexible glottal excitations providing a high degree of individuality and naturalness.
For example, specific shapes of individual glottal pulses cannot be implemented as the glottis filter is decoupled from individual excitation pulses; as the authors admit, the `use of a time-domain model of the glottal pulse makes [the approach] unsuitable for whisper decomposition'  \cite[page 892]{voice:perrotin2020}.
\item Third, explicit prosody control on a pitch pulse-by-pitch pulse basis is not allowed in the proposed approach.
\end{itemize}
Two of the synthetic voicing alternatives detailed in Section \ref{sec:experiments} overcome these limitations as they are inspired by the natural physiological process generating individual glottal pulses that excite the vocal tract.

\subsubsection{Data-driven approaches}
\label{sec:whisper2natspeech_DD}

An overview of data-driven approaches for the conversion of whispered speech to natural speech is provided in  \cite{voice:marco2022}.
Typically, data-driven approaches do not use explicit domain knowledge of the human speech production system. Instead, they make use of a specific machine learning (ML) architecture, or network, in order to project, according to a training algorithm and an optimization procedure, input speech parameters, or input data, into output speech parameters, or output data.
ML architectures include Deep Neural Networks, Generative Adversarial Networks, Autoencoders, Bidirectional Long Short-Term Memory and Transformers \cite{voice:lian2018, voice:ghosh2018, voice:konno2016, voice:lian2019, voice:rekimoto2023, voice:tan2025, voice:yamamura2025}.

These approaches may achieve good performances but they are typically data hungry and act on multidimensional signal representations that, in most cases, are neither compact nor interpretable \cite{voice:bocklet2024}.
Complete `end-to-end' solutions are uncommon  \cite{voice:pascual2018, voice:niranjan2021}. Instead, in most cases, data-driven architectures are mainly used to perform a remapping of important spectral features from whispered speech to natural speech, which are then used by a parametric vocoder (such as STRAIGHT \cite{speech:kawahara2001} or WORLD \cite{voice:morise2016}) to perform the voice/speech synthesis.

In general, data-driven approaches require large databases of pairs of whispered speech and natural speech (which may not exist for the majority of voice patients), require time-alignment or dynamic time warping processing steps and, most importantly, do not allow explicit and independent control of individual voice characteristics such as the $f_{0}$ contour, or the idiosyncratic glottal-based signature. 
 A critical aspect is that, frequently, a non-causal operation is required, and the time-scale data-driven approaches work with involves complete words and sentences.
 This means that although data-driven whispered speech-to natural speech conversion solutions may operate in real-time, to the best of our knowledge, no data-driven solutions have been shown to comply with the desired on-the-fly operation \cite{voice:rekimoto2023}.

\section{Pitch pulse-based analysis of sustained and co-articulated vowels}
\label{sec:pitchpulse}

A shift-invariant harmonic phase model is very informative and useful in the analysis, modification and synthesis of quasi-periodic signals, such as sustained or co-articulated vowel signals, as it has been detailed in \cite{sgnps:mdpi2024}.

Our previous research with a shift-invariant phase-related feature called Normalized Relative Delay (NRD), which models the shift-invariant phase structure of harmonic signals, has been mostly based on frame-based frequen\-cy-domain processing \cite{speech:ajf_2018b, speech:silva2020, speech:ajf2020_1}. The basic assumption of frame-based frequency-domain processing is that the signal is quasi-stationary within each frame \cite{sgnps:opp2010}. Depending on the relationship between the sampling frequency ($F_S$), the number of samples in each frame (i.e., the length of a frame -$N$), and the $f_{0}$ of the periodic signal under analysis, most likely, a single frame encompasses several periods, in a non-pitch synchronous way. For example, if $F_S=22050$ Hz, if $N=1024$ samples, and if $f_{0}=110$ Hz as in a typical male voice, then, the number of periods in a frame containing 1024 samples (i.e., lasting 46.44 ms) is approximately 5. In the case of the $f_{0}$ of a female voice, which typically is one octave higher than that of a male voice, the number of periods in a frame becomes 10.

Given that frequency-domain processing delivers an average NRD vector describing all the signal periods falling in a frame, we will be analyzing if, in the case of sustained vowels, the NRD vectors extracted individually for the different periods in a frame are in good agreement with the average NRD vector that is estimated using frequency-domain processing. Second, we will investigate how this scenario compares to the case where vowels are co-articulated. Given that this scenario is more relevant in running speech, and especially important in parametric-oriented synthetic voicing of whispered speech, the insights emerging from these experiments are very important to inform  algorithms synthesizing correct and natural-sounding co-articulated vowels.

Thus, in subsection \ref{sec:PPsegm}, we present a procedure allowing to extract the NRD feature vector from each period of a sustained or co-articulated vowel signal. In subsection \ref{sec:SustCoart}, we use that procedure to obtain and analyze representative results and, in subsection \ref{sec:insights}, we summarize the main insights that are relevant to inform suitable synthesis techniques.

\subsection{Phase-oriented pitch pulse segmentation and analysis}
\label{sec:PPsegm}

The Normalized Relative Delay (NRD) of harmonic $\ell$ is defined as \cite{sgnps:mdpi2024}
\begin{equation}
\label{eq:defNRD}
{\text{NRD}}_{\ell} = \frac{\phi_{\ell}- (\ell+1) \phi_{0}}{2\pi}\;,
\end{equation}
where $\phi_{\ell}$ represents the phase of harmonic $\ell=0,1,\ldots,L-1$ at the beginning of the signal frame under analysis. In particular, $\phi_{0}$ represents the phase of the $f_{0}$. $L$ represents the total number of harmonics. For convenience, in terms of parametric modeling and interpretability, NRD is taken {\em modulo\/} 1, which means that ${\text{NRD}}_{\ell} \in [0.0, 1.0[$. As detailed in  \cite{sgnps:mdpi2024}, since NRD inherits the properties of phase, then vertical phase wrapping and vertical phase unwrapping apply to NRD feature vectors. In this context, vertical phase (un)wrapping means that it is performed along the frequency axis. Two important characteristics of NRD should be highlighted. First, by definition, ${\text {NRD}}_{0}=0$, which means that the NRD phase-related feature is intrinsically time-shift invariant. Second, the NRD is independent of the $f_{0}$ \cite{sgnps:mdpi2024}. Thus, the NRD is a very convenient parametric-oriented feature that, when combined with information characterizing the magnitudes of the different harmonics relative to the magnitude of the $f_{0}$, completely determines the waveform shape of a given periodic (i.e., harmonic) signal, independently of the time-shift, of the overall magnitude, and of the $f_{0}$ \cite{sgnps:mdpi2024}.

A very convenient way to perform segmentation of the periods of a quasi-periodic wave such as a voiced vowel, which in this context we also designate as pitch periods, is to look for the onset of the $f_{0}$, i.e., the time location where $\phi_{0}=0$ rad. This can be done by computing the DFT of each pitch period, and searching for the time location that satisfies that requirement. With that purpose in mind, let us admit that the sampling frequency is at least two orders of magnitude higher than the $f_{0}$ (such that the time resolution can be considered high, which facilitates phase analysis), and let us admit that the pitch period under analysis is P samples. Moreover, let us consider that the sampled sine wave describing harmonic $\ell$ is given by
\begin{equation}
\label{eq:singleharmonic}
x_{\ell}[n] =  A_{\ell} \sin\left( \omega_{\ell} n +\phi_{\ell}\right)\;,
\end{equation}
where $A_{\ell}$ represents the amplitude of the harmonic sine wave, $n$ is the time index, $\phi_{\ell}$ is the starting phase, and  $\omega_{\ell}=(\ell+1)\omega_{0}$, with $\omega_{0} =  \frac{2\pi}{P}$.

Computing the $P$-point DFT (as a $P$-point discretization of the Fourier Transform) of $x_{\ell}[n]$  yields
\begin{equation}
X_{\ell}[k] = \frac{A_{\ell}}{2} \frac{\sin{(\alpha P)}}{\sin{\alpha}} e^{j\theta} + \frac{A_{\ell}}{2}\frac{\sin{(\beta P)}}{\sin{\beta}} e^{j\gamma},\; \; k=0,1,\ldots,P-1 \;,
\label{eq:DFTsine}
\end{equation}
where
\begin{eqnarray}
\theta =  \phi_{\ell}-\frac{\pi}{2}+\pi\left(1-\frac{1}{P}\right)\left(\ell+1-k\right)\;,\\
\gamma =  -\phi_{\ell}+\frac{\pi}{2}-\pi\left(1-\frac{1}{P}\right)\left(\ell+1+k\right)\;,\\
\alpha = \frac{\pi}{P}(\ell+1-k)\;,\\
\beta = \frac{\pi}{P}(\ell+1+k)\; .
\label{eq:DFTsine_2}
\end{eqnarray}
In the case of the fundamental frequency, Eq. (\ref{eq:DFTsine}) reduces to
\begin{equation}
X_{0}[k] = \frac{A_{o}P}{2} \delta [k-1] e^{j\left(\phi_{0}-\frac{\pi}{2} \right)} + \frac{A_{0}P}{2} \delta [k-(P-1)] e^{-j\left(\phi_{0}-\frac{\pi}{2} \right)} \;,
\label{eq:DFTsineF0}
\end{equation}
where $\delta[\cdot]$ represents the Kronecker Delta function.

Hence, the onset of the fundamental frequency can be found by searching the $n$ index in $x_{0}[n]$ (Eq. (\ref{eq:singleharmonic}), when $\ell=0$) such that the phase of the DFT bin $k=1$ is as close as possible to $-\frac{\pi}{2}$ , i.e., $\angle X_{0}[1] \approx -\frac{\pi}{2}$. In that case, the phases of all harmonics can be found using
\begin{equation}
\phi_{\ell} = \angle X_{\ell}[1+\ell] + \frac{\pi}{2}\;.
\label{eq:DFTphases}
\end{equation}
Then, the NRD coefficients can be computed using simply ${\text{NRD}}_{\ell} = \frac{\phi_{\ell}}{2\pi}$,  $k=\ell+1 = 2,\ldots,\frac{P}{2}-1$, in case $P\geq 6$ is an even number,  or $k=\ell+1 = 2,\ldots,\frac{P-1}{2}$, in case $P\geq 5$ is an odd number. This limited range of $\ell$ is due to the fact that $X_{\ell}[k]$ is conjugate symmetric (as $x_{\ell}[n]$ is real-valued). Finally, the NRD vector can be found by unwrapping the ${\text{NRD}}_{\ell}$ coefficients {\em modulo\/} 1. 

Taking advantage of these simple analytical considerations, a practical procedure allowing to segment the pitch periods of a quasi-periodic discrete-time signal $x[n]$ representing a sustained or co-articulated vowel has been implemented that consists of the following steps:
\begin{enumerate}
\item identify manually, in an approximate way, the first and last sample of the left-most pitch period in a voiced region of interest; this sets the first estimated pitch period, as well as its period $P$;
\item find the crosscorrelation function between the estimated pitch period and $x[n+S]$, where $S$ represents a shift that varies in the range $0,\dots,2P$;
\item use the two values of $S$ that correspond to the two largest local maxima in the crosscorrelation function in order to update the period $P$ of the pitch period under consideration;
\item find the value of $S \in -\lfloor\frac{P}{2}\rfloor,\dots,\lfloor\frac{P}{2}\rfloor$ such that the $P$-point DFT of $x[n+S]$ verifies $\angle X[1] \approx -\frac{\pi}{2}$ (the $\lfloor\cdot\rfloor$ function returns the largest integer of the argument), i.e., find $S= \stackrel[s]{}{\mathrm{argmin}}\left| \angle X[1] + \frac{\pi}{2}\right|$;
\item use the value of $S$ found in order to update the first and last sample of the pitch period under consideration; finally, find all $\phi_{\ell}$ values and compute the NRD vector as explained above;
\item move the signal $x[n]$ right by P samples, return to step 2 and repeat all steps till the end of $x[n]$ is reached.
\end{enumerate}
It should be noted that albeit these steps have been designed for harmonic phase analysis on a pitch period-by-pitch period basis, they also serve the purpose of harmonic magnitude analysis, as it will be illustrated in the next subsection.

\begin{figure}[htb]
\centering
\begin{minipage}{.5\textwidth}
  \centering
  \includegraphics[width=0.93\linewidth]{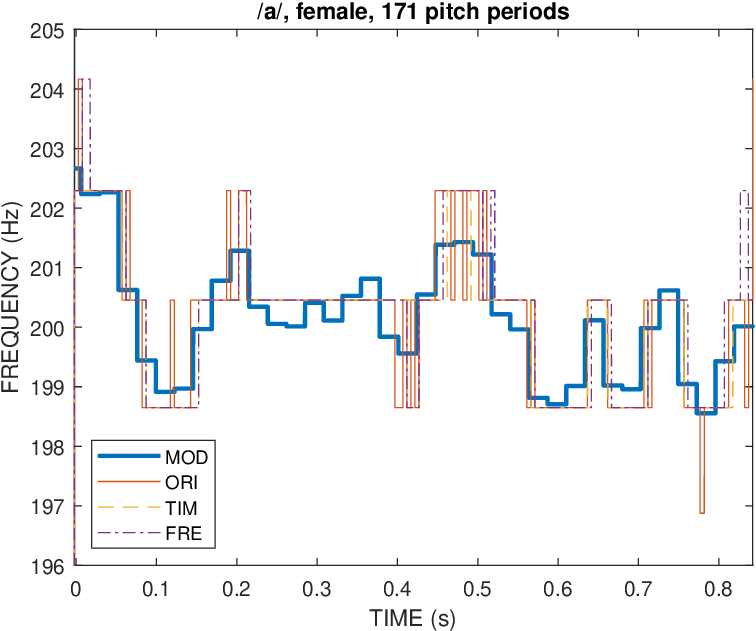}
\end{minipage}%
\begin{minipage}{.5\textwidth}
  \centering
  \includegraphics[width=0.95\linewidth]{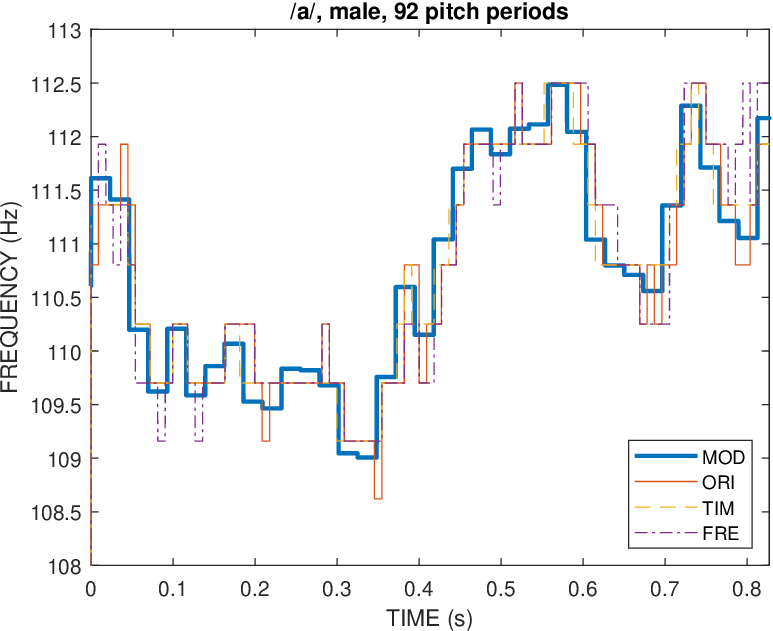}
\end{minipage}
\caption{Fundamental frequency contour of a sustained /a/ vowel uttered by a female speaker (left), and by a male speaker (right). The contours designated as `ORI' were obtained using a phase-based pitch period analysis of the original recordings, and the contours designated as `MOD' were obtained using a frame-based frequency-domain pitch estimation algorithm. The contours designated as `TIM' and `FRE' are discussed in Section \ref{sec:spectro_SV}.}
\label{fig:F0contours}
\end{figure}

In order to check the consistency between the $f_{0}$ (i.e., the reciprocal of the pitch period duration) estimated according to the above analysis procedure, and a different pitch estimation algorithm that operates in the frequency-domain \cite{speech:silvaAJF2020, speech:silva2020, sgnps:mdpi2024}, the $f_{0}$ contour of two representative sustained vowel recordings have been analyzed. The two recordings  belong to a database that is described in \cite{jesus2023}, and consist of a sustained /a/ vowel utterance produced by an adult female speaker (SPF09), and an adult male speaker (SPM05). The results are shown in Fig. \ref{fig:F0contours}. 

The $f_{0}$ contours designated as `ORI' and `MOD' were obtained using the phase-based pitch period analysis described in this subsection, and using a 512-sample frame-based frequency-domain pitch estimation algorithm \cite{speech:silvaAJF2020}, respectively. The contours designated as `TIM' and `FRE' will be addressed later on in Subsection \ref{sec:spectro_SV}. 

A few aspects should be highlighted. First, due the fact that the female pitch periods are shorter (about half) than the male pitch periods, the time resolution of the former is higher than that of the latter. Second, the staircase aspect of the pitch period-based contours just reflects the fact that the analysis procedure relies on pitch durations corresponding to an integer number of samples. Finally, given that by design the frame-based frequency-domain pitch estimation algorithm uses a frame length having a fixed number of samples (512), and estimates the average $f_{0}$ resulting from the pitch periods contained in each frame, it is not surprising that the estimated pitch contour is smoother. Overall, despite these particular aspects, the consistency between the `ORI' and `MOD' pitch contours is apparent, as intended.

\subsection{Analysis of sustained and co-articulated vowels}
\label{sec:SustCoart}

In this subsection, we use the phase-based analysis procedure described in the previous subsection to evaluate the consistency, for the same signal, between the NRD vectors obtained using frame-based frequency-domain analysis, and those obtained using pitch period-based analysis.

We take two representative sustained vowel signals corresponding to utterances of the /a/ and /i/ vowels that were produced by an adult female (SPF09), and an adult male speaker (SPM05).

\begin{figure}[htb]
\centering
\includegraphics[width=0.85\columnwidth]{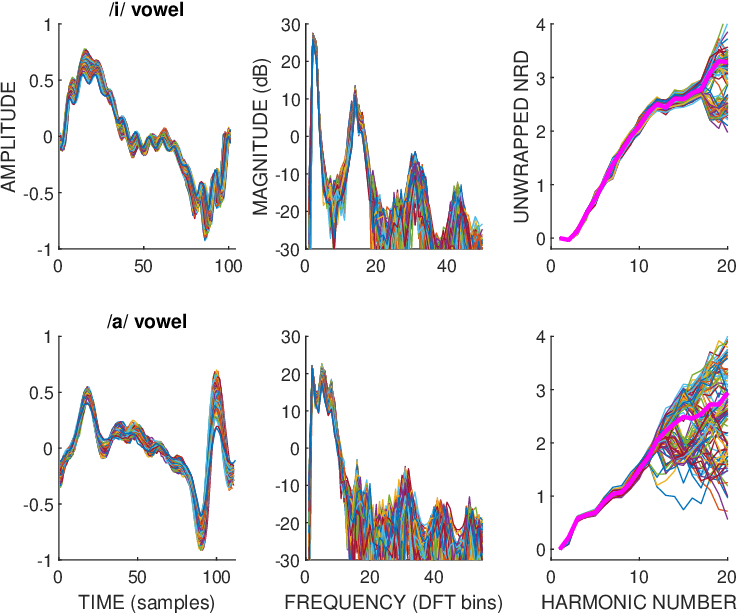}
\caption{From left to right: overlay of pitch periods found in a sustained vowel signal uttered by a female speaker, associated magnitude spectra, and associated NRD vectors. The thick magenta line represents the average of all the represented NRD vectors. The top row results from a sustained /i/ vowel signal, and the bottom row results from a sustained /a/ vowel signal.}
\label{fig:overlay}
\end{figure}

For each vowel recording pertaining to the female speaker, Fig. \ref{fig:overlay} shows an overlay of all the pitch periods that were found using the analysis procedure described in the previous subsection.
The represented /a/ vowel signal is the same as that whose $f_{0}$ contour is illustrated in Fig. \ref{fig:F0contours}.

Several relevant conclusions can be extracted from the graphical representations in Fig. \ref{fig:overlay}. First, for each vowel, the represented pitch periods are aligned and consistent, which indicates that the phase-based pitch period segmentation operates correctly. Second, as a consequence of the correct segmentation, the magnitude spectra of all the represented pitch periods are also well aligned. The exception regards the more depressed spectral regions, which is expected because those regions reflect the noise floor that is random or chaotic by nature. Finally, the NRD vectors emerging from all the represented pitch periods, and that are computed according to the analytical procedure outlined in Subsection \ref{sec:PPsegm}, are also well aligned and consistent, especially up to harmonic 17 in the case of the /i/ vowel, and up to harmonic 13 in the case of the /a/ vowel. In both cases, this is easily explained by the fact that higher-order harmonics fall in spectral regions that are depressed and, thus, more exposed to the influence of noise, which affects adversely the phase estimation process making it less robust, and less reliable  \cite{sgnps:mdpi2024}.

The NRD estimation results in Fig. \ref{fig:overlay} have been benchmarked against the NRD estimation results delivered by the frame-based frequency-domain environment that is described in \cite{sgnps:mdpi2024} and using the same /i/ and /a/ vowel signals.
It has been found that estimation results are equivalent, with negligible differences that depend on the start and end samples of the regions of the signals under analysis. This is easily explained by the fact that the graphical representations in Fig. \ref{fig:overlay} result from an analysis procedure that is pitch synchronous (as outlined in Subsection \ref{sec:PPsegm}) whereas the analysis procedure underlying the frame-based frequency-domain environment is non-pitch synchronous \cite{sgnps:mdpi2024, speech:silvaAJF2020, speech:silva2020}.
The advantage of the former over the latter, however, is that it enables studying the detailed evolution from a pitch period to the next pitch period, which is particularly important in the case of co-articulated vowels. This is addressed next.

\begin{figure}[htb]
\centering
\includegraphics[width=0.85\columnwidth]{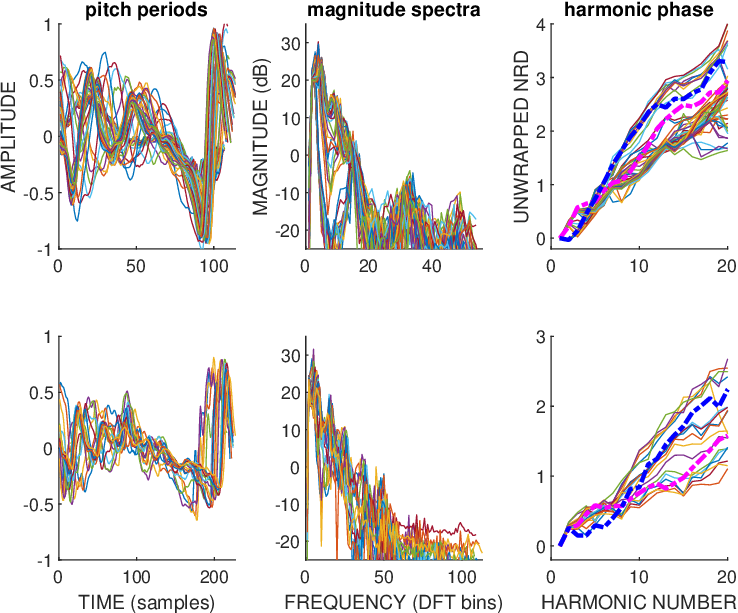}
\caption{From left to right: overlay of pitch periods found in the co-articulation of the /i/ and /a/ vowels in the Portuguese noun `Tiago', overlay of the associated magnitude spectra, and NRD vectors. The thick dashed magenta line is the reference average NRD vector due to a sustained /a/ vowel, and the thick dashed blue line is the reference average NRD vector due to a sustained /i/ vowel, both from the same speaker. The top and bottom rows represents signals produced by an adult female (SPF09) and male speaker (SPM05), respectively.}
\label{fig:overlayTiago}
\end{figure}

Figure \ref{fig:overlayTiago} illustrates an overlay of the pitch periods that were found in the co-articulation of the /i/ and /a/ vowels as uttered in the Portuguese noun `Tiago' by an adult female speaker (SPF09) and an adult male speaker (SPM05). The phase-based segmentation and analysis procedure described in Subsection \ref{sec:PPsegm} has been used to generate the different plots. Regarding the NRD vectors, the reference average NRD vector due to sustained /i/ and /a/ vowels by the same speaker are also represented to facilitate comparison.

It can be concluded that different from what was observed with sustained vowels (Fig. \ref{fig:overlay}), the overlay of pitch periods found in a  region of the signal comprising co-articulated vowels does not produce consistent plots. This is expected because each pitch period reflects a specific magnitude spectrum that varies throughout the co-articulation region, as it is also apparent from the magnitude spectrum plots in Fig. \ref{fig:overlayTiago}. However, the main objective of the analysis that motivated Fig. \ref{fig:overlayTiago} is the clarification of whether or not the NRD vectors and magnitude spectra in a co-articulation region can be looked at as an interpolation between the NRD vectors and magnitude spectra emerging from the same vowels, when produced in a sustained way by the same speaker.

Regarding the NRD vectors, and considering for illustration and discussion purposes the signals produced by the adult female speaker, it should be pointed out that in Fig. \ref{fig:overlayTiago}, the reference average NRD vectors due to sustained /i/ and /a/ vowels, are the same as those represented in Fig. \ref{fig:overlay}. Despite the fact that a perfect match is not evident --which should not be expected due to differences between phonation regimes in sustained or co-articulated vowels--, it can still be concluded that an evolution exists between NRD profiles that do not differ markedly from the NRD models emerging from the sustained vowels involved. Thus, it is fair to conclude that in what concerns NRD vectors, co-articulation regions are characterized by --and, thus, can be synthesized as-- a progressive interpolation between NRD models associated with the vowels involved, even though those models may be slightly different for the same vowels when produced by the same speaker in a sustained way, or in a word context. It is interesting to relate this conclusion with results in a previous study showing that significant differences were found between the NRD vectors pertaining to the spoken and sung versions of the same vowel produced by the same female speaker/singer \cite{speech:silva2020}. 

In Fig. \ref{fig:overlayTiago}, for both female and male speakers, up to a given harmonic (4 in the case of the female speaker, and 8 in the case of the male speaker, both cases corresponding to about 850 Hz), the average NRD vector resulting from the sustained /i/ vowel is lower than the average NRD vector resulting from the sustained /a/ vowel, while the opposite takes place above that harmonic. It is hypothesized that this may be due to the different larynx and oral tract coupling effects on the shape of the glottal excitation underlying the production of the two vowels, which could be a topic for further research.

Regarding magnitude spectra, and taking for illustrative purposes the signals produced by the female speaker considered so far in this paper, it is important to understand how the magnitude spectra in Fig. \ref{fig:overlayTiago} reflecting co-articulated vowels compare to the models of magnitude spectrum in Fig. \ref{fig:overlay} that characterize the same vowels when produced in a sustained way by the same speaker.

\begin{figure}[htb]
\includegraphics[width=0.92\columnwidth]{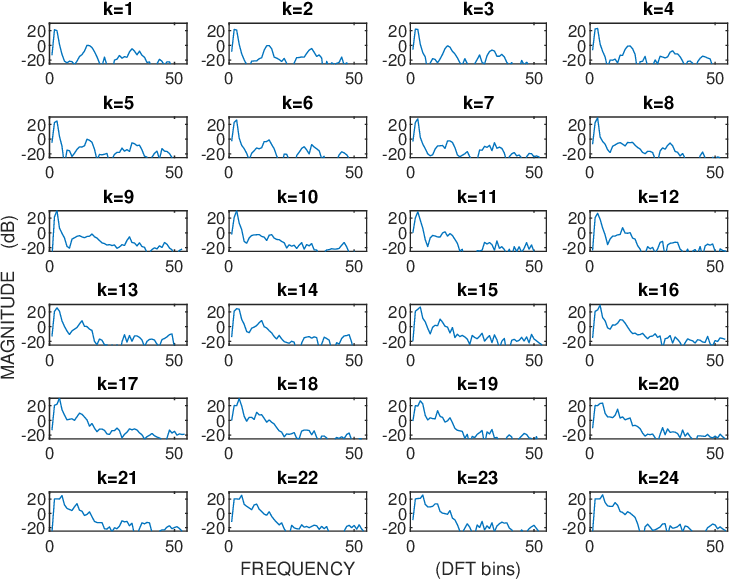}
\caption{Magnitude spectra of 24 successive pitch periods spanning the co-articulation region involving vowels /i/ and /a/ in a Portuguese word context: `Tiago'.}
\label{fig:coarticulation}
\end{figure}

The Portuguese noun `Tiago' was chosen as the word context containing a co-articulation of the /i/ and /a/ vowels. A manual spectrographic and auditory analysis of the recording of the production of this word by the same female speaker who produced the two vowels in a sustained manner, as illustrated in Fig. \ref{fig:overlay}, revealed that the co-articulation region encompasses 24 pitch periods. Figure \ref{fig:coarticulation} displays the sequence of the magnitude spectra of all the 24 individual pitch periods, the first of which corresponds to vowel /i/, and the last of which corresponds to vowel /a/.

An analysis and comparison of Fig. \ref{fig:overlay} and Fig. \ref{fig:coarticulation} leads to the following conclusions.
\begin{itemize}
\item First, regarding Fig. \ref{fig:coarticulation}, the main three formants in the magnitude spectra of the leading 7 pitch periods do not differ substantially, which suggests that the voiced region associated with the /i/ vowel does not exhibit significant co-articulation effects. Then, the subsequent 6 pitch periods (identified in the magnitude spectra as k=8 to k=13) reflect the start of the co-articulation because the third formant vanishes, the second formant slightly glides towards the first formant region, and the first formant suffers a slight broadening effect. The subsequent 7 pitch periods (identified as k=14 to k=20 in Fig. \ref{fig:coarticulation}) denote the fastest changing co-articulation region whereby the first formant moves upward in frequency, the second formant approaches even further the first formant, and a third formant emerges next to the second formant. Finally, the last 4 pitch periods (identified as k=21 to k=24 in Fig. \ref{fig:coarticulation}) reflect a stabilization of the three formants, and a stabilization of the noise floor for the highest frequencies, which denotes that the /a/ vowel associated voiced region presents no significant co-articulation effects.
\item Second, when comparing the first 7 magnitude spectra of vowel /i/ in Fig. \ref{fig:coarticulation}, and the overlay of the same-speaker /i/ vowel magnitude spectra in Fig. \ref{fig:overlay}, no major differences are observed between the associated spectral profiles. This suggest that the synthesis of vowel /i/ in a word context could be based on a spectral template inferred from a sustained vowel produced by the same speaker. However, when comparing the last 4 magnitude spectra in Fig. \ref{fig:coarticulation} that result from the production of the /a/ vowel, and the overlay of magnitude spectra in Fig. \ref{fig:overlay} for the same vowel and speaker, it can be concluded that significant differences exist between the corresponding spectral profiles, namely due to the tighter concentration of the most prominent three formants in the case of the sustained vowel production, and due to their more comparable magnitude than what is verified in the case of the co-articulated /a/ vowel production. This could be explained by the fact that a sustained articulation regime allows for the formants to be fine-tuned as a consequence of a highly steady and regulated articulatory gesture. Thus, an analysis of the results for vowel /a/ suggests that its realistic and natural synthesis in a word context, and for the same speaker, should be based on a spectral template inferred from an identical voice production regime, i.e., a word context possibly mimicking the same co-articulation sequence. 
\end{itemize}

The insights emerging from the experiments and analysis reported in this subsection are summarized in the next subsection.

\subsection{Insights regarding parametric synthesis of voicing}
\label{sec:insights}

The analysis experiments reported in the previous subsection give rise to several insights concerning the parametric synthesis of artificial voicing that is required in the restoration of natural speech from whispered speech.

\begin{enumerate}
\item The phase-based pitch period segmentation procedure described in Subsection \ref{sec:PPsegm} is appropriate to study the evolution of the $f_{0}$, harmonic phase structure, and harmonic magnitude structure of sustained or co-articulated vowel signals, on a pitch period-by-pitch period basis. Thus, phase-based pitch period segmentation and analysis is a useful tool assessing the operational correctness and sound quality delivered by algorithms implementing synthetic voicing of whispered vowels.
\item The harmonic phase/magnitude structure of vowels in a co-articulation context may differ from the harmonic phase/magnitude structure of the same vowels when produced in a sustained way by the same speaker, this appears to be true in the case of vowels whose formants are not far apart and appear instead as clusters of formants, as in the case of the F1 and F2 formants that characterize the typical spectral envelope of the /a/ vowel. This may be due to the fact that articulation gestures in co-articulation regions are different from those characterizing sustained vowels produced by the same speaker, namely in terms of the coupling between the larynx and the vocal tract, which impacts on the vocal fold operation, i.e., on the shape of the glottal pulse. Thus, the spectral templates used by parametric-oriented algorithms for the restoration of natural speech from whispered speech should be based as much as possible on templates inferred from natural co-articulation regions in running speech.
\item Independently of the mismatch identified in the previous point, the pitch period-based evolution of the harmonic phase/magnitude structure of co-articulated vowels can be looked at as a progressive interpolation between the natural --and idiosyncratic-- spectral profiles characterizing those vowels in a running speech context. Thus, provided that those spectral profiles are realistically modeled by suitable spectral templates, an algorithm for the restoration of natural speech from whispered speech may synthesize entire co-articulation regions as a result of the progressive pitch period-based interpolation between those reference spectral templates.
\end{enumerate}

\section{Signal processing framework for synthetic voicing}
\label{sec:experiments}

In this section, we focus on a specific and parametric-oriented signal processing module that is responsible for generating synthetic voicing and that plays a key role in a general signal processing framework converting whispered speech into voiced speech, not only in real-time, but also on-the-fly. 
The careful segmentation of those regions in whispered speech that should have synthetic voicing implanted is critical to the success of the whole voice restoration process. Because careful and precise segmentation elicits challenges of its own  \cite{voice:goncalo2023, voice:costa2021}, and involves sophisticated processing, it is out of the scope of this paper.

Specifically, in this section, we describe and compare, from an operational point of view, three signal processing alternatives allowing the reconstruction of synthetic voicing that is mixed with the whispered speech in selected regions.

The relative quality of the three signal processing alternatives will be evaluated in Section \ref{sec:listening} taking as ground truth the original recordings of sustained and co-articulated vowels in word context, and using reconstructed versions of those recordings after replacement of the original voiced regions by synthetic voicing created by each one of the discussed synthetic voicing alternatives.

Thus, in Subsection \ref{sec:FDanal}, we briefly address the signal analysis extracting relevant parametric information from a quasi-periodic waveform and that is necessary to drive the three synthetic voicing alternatives. These alternatives are addressed in Subsections \ref{sec:FDsynt}, \ref{sec:TDsyntConPP}, and \ref{sec:TDsyntOAPP}, in a perspective that highlights their main differences in terms of preferred domain (frequency or time), and signal reconstruction approaches.

\subsection{Frequency-domain parametric analysis framework}
\label{sec:FDanal}

\begin{figure}[htb]
\begin{center}
  \input{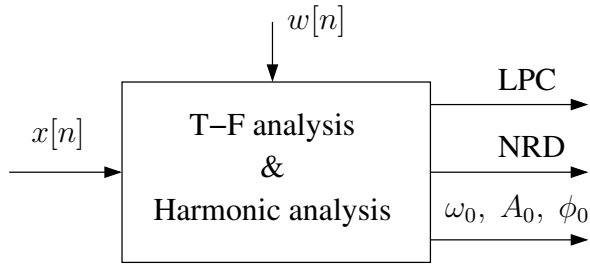}
\end{center}
\caption{Simplified frequency-domain analysis framework that takes as input a quasi-periodic waveform ($x[n]$) and yields a parametric representation allowing its reconstruction: the angular fundamental frequency ($\omega_0$), its magnitude ($A_0$) and phase ($\phi_0$), and a shift-invariant harmonic phase model (NRD) and magnitude model (LPC).}
\label{fig:gen_enc}
\end{figure}

Figure \ref{fig:gen_enc} represents a simplified block diagram of the general frequency-domain parametric analysis framework that we use in this paper.
This analysis framework has been detailed in \cite{speech:ajf2020_1, speech:silva2020} and extracts parametric information from a quasi-periodic waveform (e.g., a sustained or co-articulated vowel represented by $x[n]$) allowing its reconstruction: frequency, magnitude and phase of the fundamental frequency ($\omega_0$, $A_0$, $\phi_0$), a shift-invariant harmonic phase model (NRD), and a harmonic magnitude model (inferred from an LPC model).

The analysis framework involves two main processing stages that operate in a frame-based manner: a front-end analysis, and a parametric analysis, as described next.
\begin{itemize}
\item The front-end consists of a frame-based Time-Frequency (T-F) transformation by means of an Odd-frequency Discrete Fourier Transform (ODFT) whose sub-bands are determined by a window function ($w[n]$) that corresponds to the square root of a shifted Hanning window \cite{sgnps:mdpi2023, sgnps:vaidya93}; fixed segmentation using $N=1024$ samples, and 50\% overlap between adjacent frames are implemented.
\item The parametric analysis consists of accurate harmonic frequency and pitch estimation  \cite{sgnps:mdpi2023, speech:silvaAJF2020},  phase and magnitude estimation \cite{sgnps:mdpi2024, sgnps:ajfmoh01_a}, harmonic phase modeling using NRD \cite{sgnps:mdpi2024}, and harmonic magnitude modeling by means of LPC analysis \cite{sgnps:hayes96}.
\end{itemize}
The sampling frequency of the recordings considered in this paper is 22050 Hz, and the sample resolution is 16 bits.

\subsection{Frequency-domain parametric synthesis of voicing}
\label{sec:FDsynt}

\begin{figure}[htb]
\begin{center}
  \input{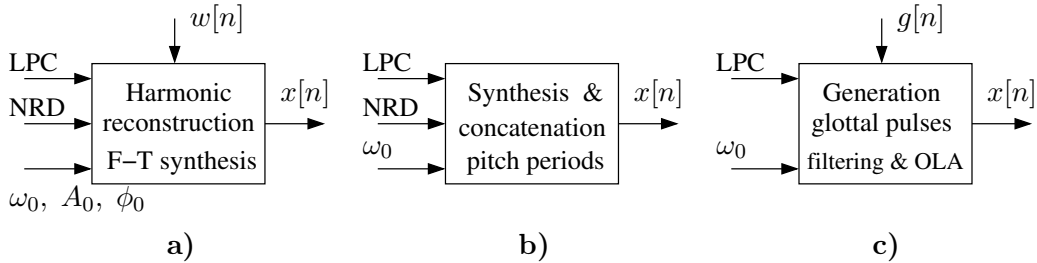}
\end{center}
\vspace{-0.5cm}
{\small\bf\hspace{20mm} a) \hspace{40mm} b) \hspace{40mm} c)}
\caption{Three parametric-oriented alternatives to synthesize artificial voicing: {\small\bf a)} entirely frequency-domain synthesis as discussed in Subsection \ref{sec:FDsynt}; {\small\bf b)} combined frequency and time-domain synthesis as discussed in Subsection \ref{sec:TDsyntConPP}; and {\small\bf c)} physiologically inspired time-domain synthesis as discussed in Subsection \ref{sec:TDsyntOAPP}.}
\label{fig:mul_dec}
\end{figure}

The first parametric-based approach to the synthesis of artificial voicing that we consider in this paper is entirely frequency-domain, and is illustrated in Fig. \ref{fig:mul_dec}a. It corresponds to the same synthesis procedure that is reported in \cite{speech:ajf2020_1}, and consists of the reverse steps taken in the analysis. A complex, discrete-frequency, spectrum is first reconstructed on the assumption that $N \geq 3L$, where $L$ is the number of harmonics in the Nyquist range. Using $\omega_{0}$, 9 ODFT bins are synthesized  for each sinusoid of the harmonic structure \cite{speech:ajf2020_1}. Magnitude information is reconstructed using $A_{0}$, the LPC model, and the accurate frequency response magnitude of window $w[n]$. Phase information is reconstructed using $\phi_{0}$, the NRD model \cite{sgnps:mdpi2024}, and the accurate frequency response phase of window $w[n]$, as explained in \cite{speech:ajf2020_1}.

\begin{figure}[htb]
\begin{center}
  \input{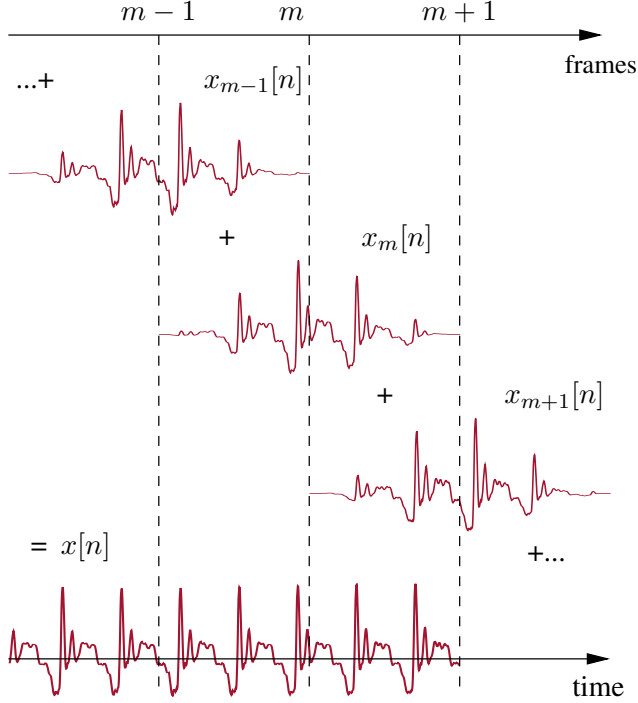}
\end{center}
\caption{ Reconstruction of $x[n]$ according to the frequency-domain synthesis approach. After being multiplied by window $w[n]$, adjacent segments including $x_{m-1}[n]$, $x_{m}[n]$, and $x_{m+1}[n]$ are overlapped and added yielding $x[n]$.}
\label{fig:recM05FREQ}
\end{figure}

Following the reconstruction, in the ODFT domain, of the harmonic structure in each frame, a Frequency-Time (F-T) transformation takes place by means of an inverse ODFT. The resulting time-domain segment is subsequently multiplied by window $w[n]$, and overlapped and added to the two neighboring segments (with a 50\% overlap) to create the final time-domain signal, i.e,
\[
x[n]=\sum_{m} x_{m}\left[n-m\frac{N}{2}\right]\;.
\]
 This final reconstruction step is illustrated in Fig. \ref{fig:recM05FREQ} that represents actual data taken from our processing environment when a sustained /a/ vowel signal produced by a male speaker (SPM05) is used. In Fig. \ref{fig:recM05FREQ}, the windowed segments are represented as $x_{m}[n]$, where $m$ denotes the frame index.
The main advantages of this method are: its computational efficiency, the signal quality resulting from the inherent match between analysis and synthesis, and the flexibility in controlling the magnitude and phase harmonic structures independently.

Regarding the inherent match between analysis and synthesis, it should be noted that the cascade of the frequency-domain analysis depicted in Fig. \ref{fig:gen_enc}, and the frequency-domain synthesis just described, when framed in the overlap-add scheme illustrated in Fig. \ref{fig:recM05FREQ} and relying on a window satisfying specific requirements --as the one used in this paper--, and in the absence of spectral modification, leads to a perfect reconstructing analysis-synthesis framework \cite{sgnps:vaidya93, filtdsg:malvar92}, as it is commonly used in the context of audio coding/compression.

A limitation of this method has to do with the basic assumption underlying its frequency-domain, frame-based, nature and operation: the quasi-stationary of the signal within the boundaries of each frame that in our experiments spans 1024/22050=46.4 ms. This assumption is fairly realistic as long as the $f_{0}$ and magnitude and phase structure of the individual pitch periods falling within those boundaries do not change significantly. As co-articulation regions may elicit challenges from this point of view, a signal reconstruction oriented to the synthesis of individual pitch periods may be beneficial. This is the main motivation regarding the methods that are described in the next two subsections.

\subsection{Combined frequency and time-domain synthesis of voicing}
\label{sec:TDsyntConPP}

A simplified block diagram in Fig. \ref{fig:mul_dec}b illustrates the combined frequency and time-domain synthesis of voicing. Pitch periods are individually synthesized using
\begin{equation}
\label{eq:pitchperiods}
x_{p}[n]  =   \sum_{\ell=0}^{L-1} A_{\ell,p} \sin\left( n\left(\ell+1\right)\omega_{0,p} + 2\pi{\text{NRD}}_{\ell,p}\right),\;\; n=0,1,\ldots,P_{p}-1\;,
\end{equation}
where $\omega_{0,p}=\frac{2\pi}{P_p}$, and $P_p$ is the number of samples of the pitch period having index $p$.

It should be emphasized that, in Eq. (\ref{eq:pitchperiods}), the main synthesis parameters ($\omega_{0,p}$, $A_{\ell,p}$, and ${\text{NRD}}_{\ell,p}$) result from an interpolation of the frequency, magnitude (which is derived from the LPC model), and NRD information that is available from (and updated at) the two frame boundaries surrounding the pitch period indexed as $p$. The interpolation is calibrated as a function of the relative position of the pitch period with respect to those boundaries.

\begin{figure}[htb]
\begin{center}
  \input{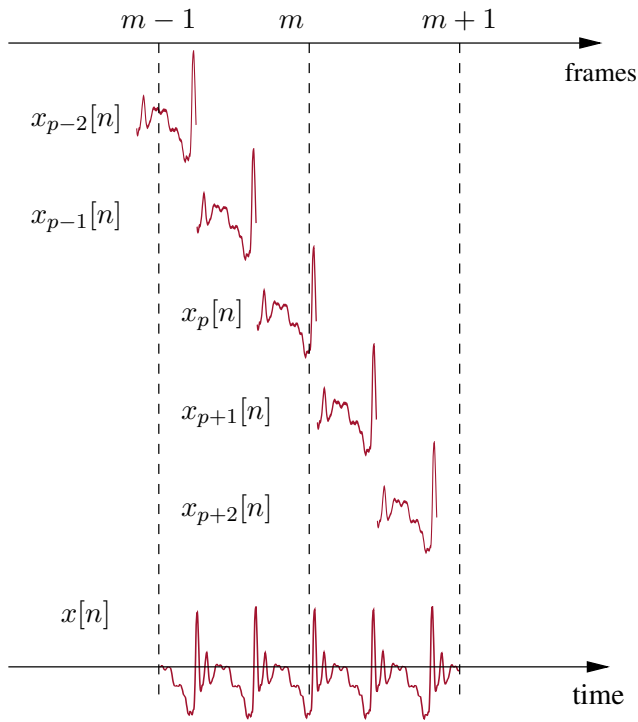}
\end{center}
\caption{Reconstruction of $x[n]$ according to the combined frequency and time-domain synthesis alternative. Pitch periods are individually shaped in frequency, synthesized and then concatenated in the time domain.}
\label{fig:recM05TIME}
\end{figure}

As illustrated in Fig. \ref{fig:recM05TIME}, after synthesis, the individual pitch periods are concatenated to give rise to the final signal represented by $x[n]$,
i.e,
\[
x[n]=\sum_{p} x_{p}\left[n - n_p\right]\;,
\]
where $n_p$ accumulates past pitch periods, i.e., $n_p=\sum_{k=-\infty}^{p-1}P_{k}$.
In particular, the time location of the individual pitch periods is determined by the $f_{0}$ contour that is available at frame boundaries but is not constrained by the frame boundaries, as Fig. \ref{fig:recM05TIME} illustrates. 

In our implementation, we adopt a simple post-processing strategy when two periods are concatenated: they are circularly extended by 4 samples on each side, and a piecewise linear fade-in and fade-out rule is enforced in the small region where the two adjacent (and circularly extended) pitch periods overlap. This post-processing is meant to mitigate a minor sound quality problem (in the form of a faint `impulsive noise and subtle pitch alterations') that was detected when this synthesis method was first implemented and tested, as reported in \cite{speech:ajf2020_1, pc:ajfaes16}. In fact, despite Eq. (\ref{eq:pitchperiods}) ensuring that the DC of each pitch period is zero, differences in length and spectral envelope between adjacent periods may give rise to slightly audible signal discontinuities if pitch periods are simply concatenated.

When compared to the method described in the previous subsection, the combined frequency and time-domain synthesis of voicing allows for a flexible synthesis of individual pitch periods, which makes the time and spectral evolution of a sequence of pitch periods not only more under control and smoother from a signal processing standpoint, but also potentially more naturally sounding from a perceptual point of view because it mimics better the natural voice production mechanism. A disadvantage is that it is not as computationally efficient as the previous (frequency-domain) method. In addition, algorithm complexity is slightly aggravated because pitch periods and frame boundaries are not synchronized, as Fig. \ref{fig:recM05TIME} shows.

As a final note, it should be pointed out that the shape of the waveforms reconstructed by the two methods discussed so far are expected to be similar given that they rely on the exact same reference magnitude and phase parameters, as the $x[n]$ plots in Fig. \ref{fig:recM05FREQ} and Fig. \ref{fig:recM05TIME} representing actual data suggest. The most visually noticeable difference between the outputs of the two methods is a sub-period relative shift at the beginning of the stream of pitch periods because the starting phase of the $f_{0}$ is not explicitly used by the method described in this subsection. This specific aspect has no relevant perceptual implications (i.e., it is not audible). However, subtle audible differences may exist due to the fact that the signal processing techniques --and associated frequency and time support-- characterizing the two synthesis methods described so far are different.

The method described in the next subsection generates waveforms whose shape is different from the shape of the waveforms synthesized by the two methods discussed so far, despite having the same spectral envelope (i.e., the same magnitude spectrum).

\subsection{Physiologically inspired time-domain parametric synthesis of voicing}
\label{sec:TDsyntOAPP}

\begin{figure}[htb]
\begin{center}
  \input{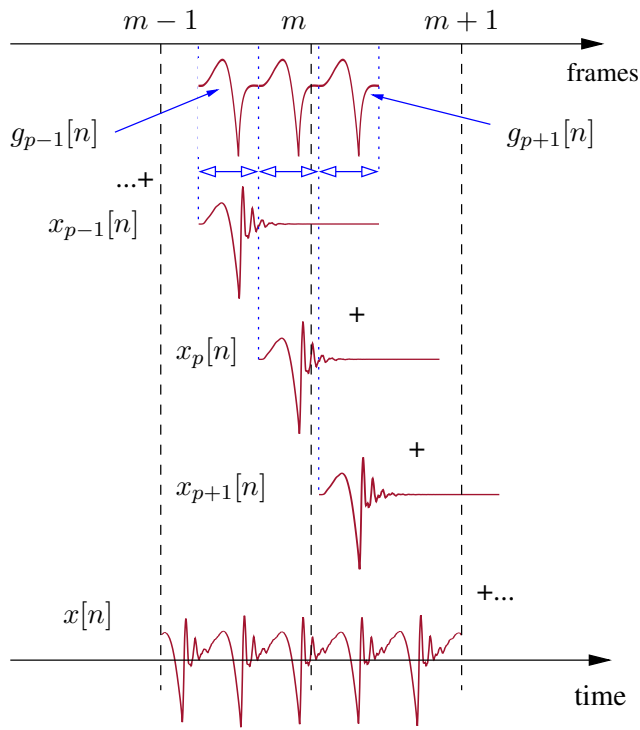}
\end{center}
\caption{Reconstruction of $x[n]$ according to the physiologically inspired time-domain synthesis alternative. Glottal excitation periods are synthesized individually, are filtered according to a filter modeling an individual spectral envelope, and are overlapped and added to yield the final time-domain signal $x[n]$.}
\label{fig:recM05glottal}
\end{figure}

Figure \ref{fig:mul_dec}c represents a simplified block diagram of the third alternative considered in this paper implementing parametric synthesis of voicing. Figure \ref{fig:recM05glottal} depicts schematically its main signal processing steps.
The synthesis is entirely time-domain and is based on the physiological process underlying the natural voice production mechanism:
\begin{itemize}
\item individual pitch pulses representing the glottal excitation are first synthesized using a synthesis procedure relying on parametric information describing the idealized Liljencrants-Fant model, as explained in \cite{speech:ajf_2018b}; the individual glottal excitation pulses are represented in Fig. \ref{fig:recM05glottal} as $g_{p}[n]$, where $p$ denotes the glottal pulse index; as in the previous method, the number of samples of the $p$-th glottal pulse period is $P_p$ and is delimited in Fig. \ref{fig:recM05glottal} by a horizontal line having double arrows; also, as in the previous method, the time location of the individual pulses is determined by the angular fundamental frequency ($\omega_{0}$) contour that is available at frame boundaries, although it is not constrained by those boundaries;
\item each glottal excitation pulse is filtered by a dedicated vocal tract filter ($h_{p}[n]$) modeling the spectral envelope resulting from the vocal tract shape during the specific time span of that pulse, and taking into consideration the spectral tilt characterizing the spectral profile of the glottal excitation pulse \cite{voice:ajfaes18, speech:AJF2024},  i.e.,
\[
x_{p}[n] = g_{p}[n] \ast h_{p}[n] = \sum_{k} g_{p}[k] h_{p}[n-k]\; ;
\]
illustrative filtered pulses ($x_{p}[n]$) are represented in Fig. \ref{fig:recM05glottal} using actual data from a sustained /a/ vowel signal produced by a male speaker (SPM05);
\item as in the synthesis alternative described in the previous subsection, the filtered pulses are overlapped and added to reconstruct the final output signal $x[n]$.
\end{itemize}
Although this synthesis alternative shares some of the specificities of the one discussed in the previous subsection, such as flexible, autonomous, and physiologically inspired synthesis of individual pitch periods, two important differences exist.

First, as the block diagram in Fig. \ref{fig:mul_dec}c denotes, the synthesis alternative discussed in this section does not use external NRD information, which means that the harmonic phase structure is embedded in the synthesis of the glottal excitation. This is motivated by the convenience of using All-Pole (AP) filters \cite{sgnps:hayes96} --namely in terms of computational efficiency and ability to model vocal tract resonances--, and by the results of a recent study on the phase distortion caused by AP filters \cite{speech:AJF2024}. Those results arise from the synthesis of sustained vowel signals and show that although the objective impact of a vocal tract filter implemented as an AP filter in changing the harmonic phase structure of a plausible glottal excitation is measurable and significant, its perceptual impact is little and mostly not audible.

Second, as the graphical representations of $x_{p}[n]$ in Fig. \ref{fig:recM05glottal} suggest, the linear convolution between $g_{p}[n]$ and $h_{p}[n]$ gives rise to a local output ($x_{p}[n]$) whose relevant part may have a duration significantly exceeding the time span ($P_p$) of the involved glottal excitation pulse. In our processing framework, we adopt a conservative duration equal to three times the length of the local period ($P_p$).

Thus, linear and time-shift invariant filtering is locally preserved by overlapping and adding the local output signals resulting from the specific filtering of individual glottal excitation pulses. It should be pointed out that this is different from filtering a sequence of pulses using a filter whose impulse response changes through time to reflect the changing spectral influence of the vocal tract filter. In this case, the filtering is linear, but strictly not time-shift invariant, which may introduce --potentially audible-- artifacts of its own. 

Two main advantages of the physiologically inspired time-domain parametric synthesis of voicing should be emphasized. On the one hand, each individual glottal excitation pulse preserves the same waveform shape, independently of its period, because it is individually synthesized taking this period in consideration and taking that concern (waveform shape) into consideration. On the other hand, contrary to the two previously discussed voicing synthesis alternatives where the glottal excitation is embedded in the spectral representation of the signals, but not factorizable, in the physiologically inspired time-domain parametric synthesis of voicing the glottal excitation is explicitly accessible and decoupled from the vocal tract influence. This means that, in this case, an additional flexibility exists to tailor the glottal excitation according to the idiosyncrasies and operation of the vocal folds and larynx of a specific speaker, which helps in giving the synthetic voicing the desired sound signature of that speaker, or a specific phonation type, e.g., modal, tense, or breathy.

\section{Experiments, results and discussion}
\label{sec:expresdisc}

The objective and perceptual impact of the three alternatives for synthesizing artificial voicing, as discussed in Subsections \ref{sec:FDsynt}, \ref{sec:TDsyntConPP}, \ref{sec:TDsyntOAPP}, and illustrated in Fig. \ref{fig:mul_dec}, have been assessed using recordings of sustained /i/ and /a/ vowels produced by an adult female (SPF09), and an adult male speaker (SPM05), as well as recordings of the same vowels co-articulated in a word context corresponding to the voiced realization of the Portuguese noun `Tiago' by the same speakers. Another word corresponding to the Portuguese verb `Saiu' has been added to enrich the perceptual experiments. 
In both cases (sustained and co-articulated vowels), background noise has also been synthesized and added to the signal 25 dB below the spectral envelope of the harmonics. The objective of this processing was to mimic the natural noise floor that exists in a typical vowel realization by a human speaker.

In Subsections \ref{sec:spectro_SV} and \ref{sec:spectro_coartV} we discuss a comparative analysis of the three technical alternatives using spectrograms. In Subsection \ref{sec:listening} we describe the listening tests that were organized to assess the relative perceptual impact of the three technical alternatives, and we discuss the associated results.

\subsection{Experiments with sustained vowels}
\label{sec:spectro_SV}

\begin{figure}[htb]
\centering
\includegraphics[width=0.85\columnwidth]{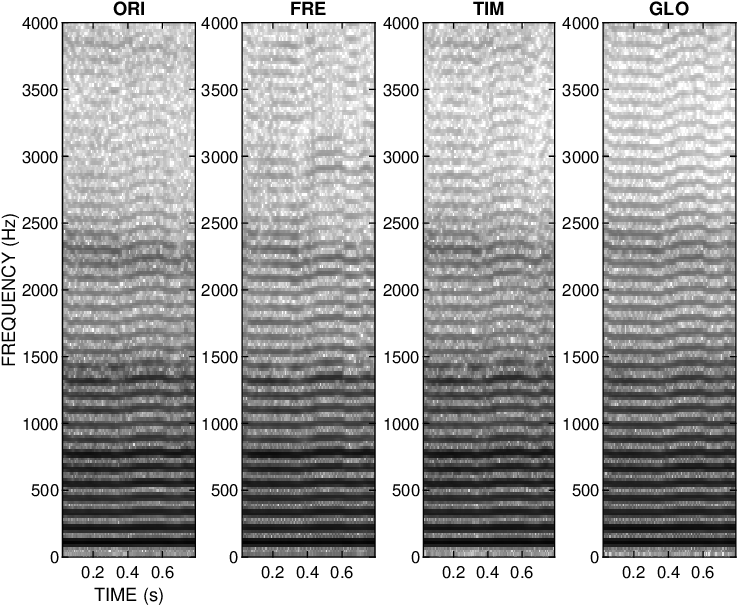}
\caption{Spectrograms of the original (ORI) and synthetic versions (FRE, TIM, GLO) of a sustained /a/ vowel produced by a male speaker. Darker colors denote higher Power Spectral Densities.}
\label{fig:SPM05_04_01_a_3}
\end{figure}

The synthetic vowels were obtained using the frequency-domain (FRE), combined frequency and time-domain (TIM), and physiologically inspired time-domain (GLO) algorithms.
As a representative example of the sustained vowels, Fig. \ref{fig:SPM05_04_01_a_3} shows the spectrograms of the original (plot identified as `ORI') and three synthetic versions (plots identified as `FRE', `TIM', and `GLO') of a sustained /a/ vowel produced by an adult male speaker (SPM05). This vowel realization is the same as that underlying the plots on the right-hand side of Fig. \ref{fig:F0contours}.

The spectrograms motivate the following remarks.
\begin{itemize}
\item Four main formant frequencies are easily identified at around 200 Hz, 700 Hz, 1400 Hz, and 2300 Hz, and are consistent across spectrograms. This indicates that their algorithmic realization is correct independently of the synthesis method.
\item Of all the spectrograms representated in Fig. \ref{fig:SPM05_04_01_a_3}, the one resulting from the physiologically inspired time-domain synthesis (and identified as `GLO') is the one whose harmonic structure integrity is well defined up to 4000 Hz, even supplanting in this regard the harmonic structure integrity of the original signal. This reveals that a merit of the GLO approach, which is based on the linear convolution between each individual glottal excitation pulse and the associated filter materializing the spectral influence of the vocal tract filter during the time span of that pulse, is that it can be used as a technique repairing damaged harmonic structures due, for example, to a dysphonia condition.
\item The $f_{0}$ contour, which is especially noticeable for higher order harmonics, is clearly in line with the $f_{0}$ contour in Fig. \ref{fig:F0contours} regarding the same vowel and speaker, which supports the results in that figure. This has motivated an additional experiment that is described next.
\end{itemize}
The $f_{0}$ contours of the synthetic sustained vowels have been analyzed using the phase-based pitch analysis algorithm that is described in Subsection \ref{sec:PPsegm}. Representative results regarding the sustained /a/ vowel are shown in Fig. \ref{fig:F0contours}. The contours identified in this figure as `FRE' and `TIM' correspond to the outputs of the frequency-domain synthesis algorithm (Subsection \ref{sec:FDsynt}) and the time-domain synthesis algorithm (Subsection \ref{sec:TDsyntConPP}), respectively. The output of the physiologically inspired time-domain synthesis algorithm (Subsection \ref{sec:TDsyntOAPP}) is not shown because it overlaps the displayed `TIM' contour. It can be concluded that the different plots are quite consistent (occasional differences are due to episodic differences of 1 sample in the estimation of the period length), which confirms that the proposed phase-based pitch analysis algorithm can also be used to monitor the correct operation of synthetic voicing algorithms.

\subsection{Experiments with co-articulated vowels}
\label{sec:spectro_coartV}

Two Portuguese words containing co-articulated vowels were selected to be included in the synthesis experiments. The two words, `Tiago' and `Saiu', were produced by the female (SPF09) and male (SPM05) speakers considered in this paper.

As indicated at the beginning of Section \ref{sec:experiments}, the synthetic versions of the words were obtained by completely replacing the original voiced regions by synthetic voicing created by each one of the synthesis techniques discussed in Subsections \ref{sec:FDsynt}, \ref{sec:TDsyntConPP}, and \ref{sec:TDsyntOAPP}.

\begin{figure}[htb]
\centering
\includegraphics[width=0.85\columnwidth]{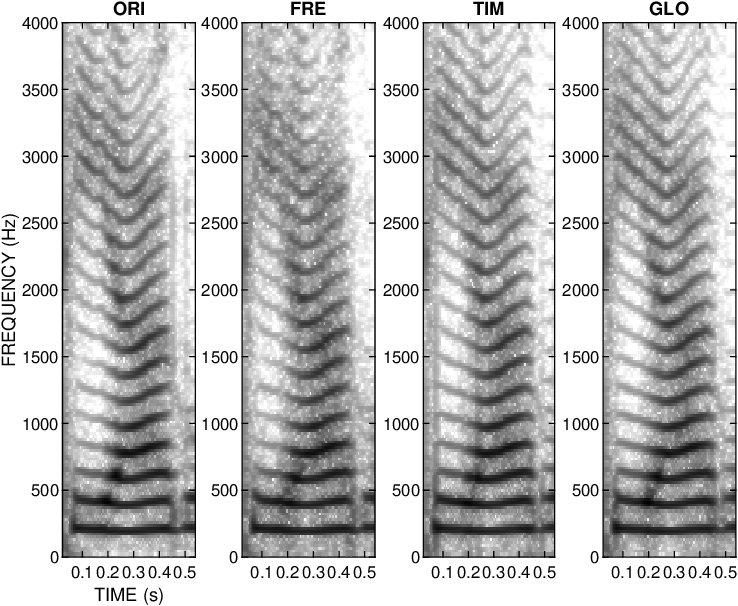}
\caption{Spectrograms of the original (ORI) and synthetic versions (FRE, TIM, GLO) of the word  `Tiago' produced by a female speaker. Darker colors denote higher Power Spectral Densities.}
\label{fig:SPF09_47_01_01_Tiago_3}
\end{figure}

Fig. \ref{fig:SPF09_47_01_01_Tiago_3} shows the spectrograms of the original and synthetic versions of the relevant region containing the co-articulated vowels /i-a/ in the `Tiago' word context produced by the female speaker (SPF09). As described above, the synthetic co-articulated vowels were obtained using the frequency-domain (FRE), combined frequency and time-domain (TIM), and physiologically inspired time-domain (GLO) algorithms.
The spectrograms motivate the following remarks.
\begin{itemize}
\item The $f_{0}$ contour that initiates at around 200 Hz
at the start of the /i/ vowel is consistent across spectrograms, although higher order harmonics exhibit a better continuity in time for the TIM and GLO spectrograms relative to the ORI and FRE cases. This confirms that time-domain synthesis has the ability to preserve harmonic structure integrity in the synthetic signals.
\item  The co-articulation region is clearly and consistently manifested in all spectrograms starting with a wide separation between the F1 and F2 formant frequencies (in the form of a gap existing between 600 Hz and 2500 Hz) that typically characterizes the /i/ vowel, and which progressively closes down between 0.05 s and 0.25 s to give rise to the realization of the /a/ vowel; during this time, the $f_{0}$ decreases, as expected, because of the higher intrinsic $f_{0}$ of the /i/ vowel, when compared to the /a/ vowel.
\item The transition between the voiced consonant /g/ and the /o/ vowel is clear and consistent in all spectrograms between 0.4 s and 0.5 s.
\end{itemize}
Overall, a spectrogram-based analysis suggests that, from a signal processing point of view, not only the same phonemes in the different synthetic realizations possess comparable spectral properties, but also the spectral dynamics in the co-articulation region is correct, independently of the synthesis algorithm. 

\begin{figure}[htb]
\centering
\includegraphics[width=0.85\columnwidth]{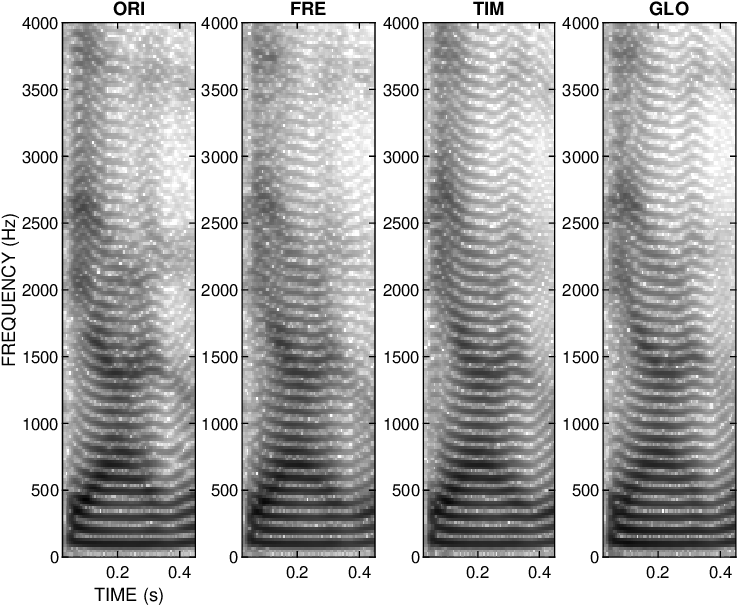}
\caption{Spectrograms of the original (ORI) and synthetic versions (FRE, TIM, GLO) of the word  `Tiago' produced by a male speaker. Darker colors denote higher Power Spectral Densities.}
\label{fig:SPM05_47_01_01_Tiago_3}
\end{figure}

Similar remarks can be made with respect to Fig. \ref{fig:SPM05_47_01_01_Tiago_3} that shows the same sequence of spectrograms representing the
word `Tiago'  produced by the speaker (SPM05).
The spectrograms confirm that time-domain synthesis (spectrograms identified as TIM and GLO) has the ability to preserve harmonic structure integrity in the synthetic signals. In addition, spectrograms suggest that a better formant continuity is achieved in the two cases involving time-domain synthesis because the trajectories of the formants in the spectral region above 1500 Hz are smoother and uninterrupted. 

\begin{figure}[htb]
\centering
\includegraphics[width=0.85\columnwidth]{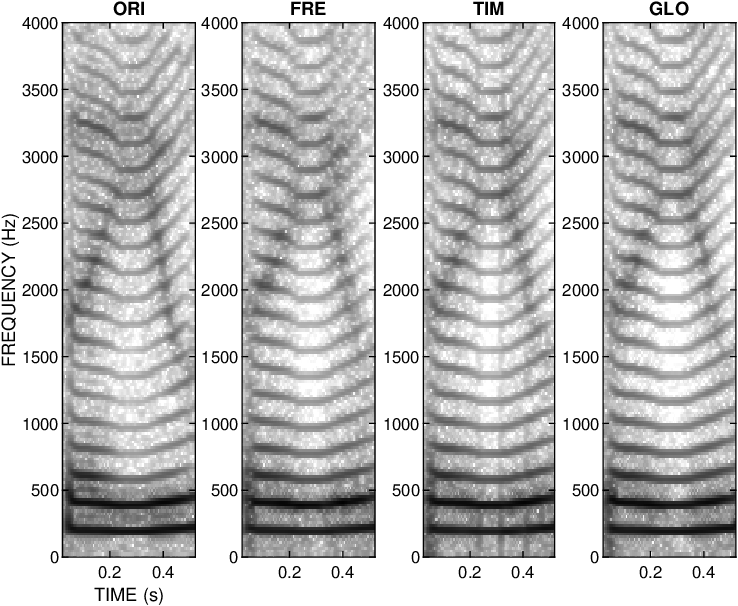}
\caption{Spectrograms of the original (ORI) and synthetic versions (FRE, TIM, GLO) of the relevant voiced region of the word  `Saiu' produced by a female speaker and highlighting the co-articulation of the three vowels involved. Darker colors denote higher PSD levels.}
\label{fig:SPF09_43_01_01_saiu_3}
\end{figure}

\begin{figure}[htb]
\centering
\includegraphics[width=0.85\columnwidth]{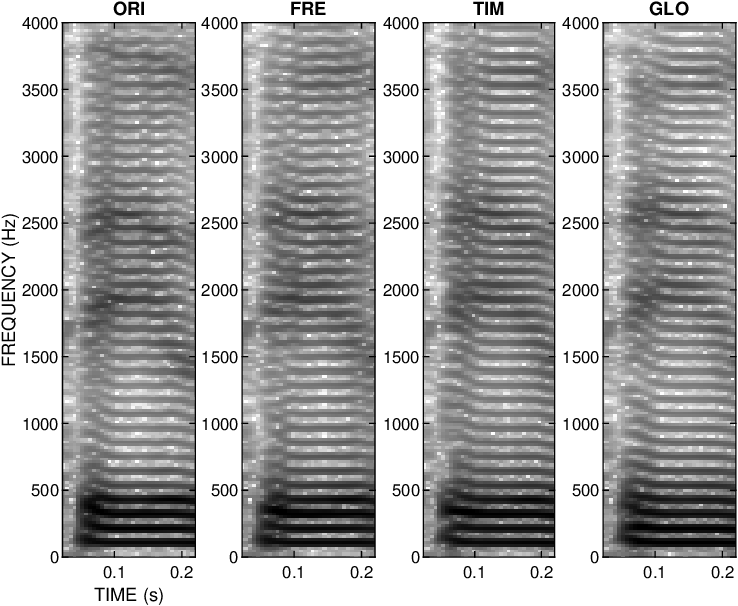}
\caption{Spectrograms of the original (ORI) and synthetic versions (FRE, TIM, GLO) of the relevant voiced region of the word  `Saiu' produced by a male speaker and highlighting the co-articulation of the three vowels involved. Darker colors denote higher PSD levels.}
\label{fig:SPM05_43_01_saiu_3}
\end{figure}

Figures \ref{fig:SPF09_43_01_01_saiu_3} and \ref{fig:SPM05_43_01_saiu_3} show the same sequence of spectrograms concerning the relevant voiced region in the realization of the word `Saiu' by an adult female (SPF09) and male (SPM05) speaker, respectively.
Given that the word realization by the male speaker is shorter --about half of the time-- than in the case of the same-word realization by the female speaker, the $f_{0}$ contour is steadier, and the formant trajectories are less evident in the spectrograms showing the word realization by the male speaker.
Still, we could observe that, in both cases and taking the original signal as a reference, the $f_{0}$ contour, as well as the formant frequency trajectories, appear to be correctly reconstructed in the synthetic versions of the word. These remarks corroborate the results previously presented for the word `Tiago'.

Thus, from an objective point of view, at least from what can be inferred using spectrograms, there are no clear signs anticipating major positive or negative perceptual impacts due to the three different voicing reconstruction techniques discussed in Subsections \ref{sec:FDsynt}, \ref{sec:TDsyntConPP}, and \ref{sec:TDsyntOAPP}. The goal of the listening experiments described and discussed in the next subsection is precisely to assess those perceptual --or subjective-- impacts.

\subsection{Listening experiments}
\label{sec:listening}

Subjective listening tests were organized and conducted in order to assess the perceptual impact of each one of the three synthetic voicing techniques discussed in Subsections \ref{sec:FDsynt}, \ref{sec:TDsyntConPP}, \ref{sec:TDsyntOAPP}. The number of original audio items included in the listening tests is eight and consist of recordings from
\begin{itemize}
\item sustained /i/ and /a/ vowels produced by an adult female (SPF09), and an adult male speaker (SPM05), totalizing 4 items;
\item co-articulated /i/ and /a/ vowels in a word context corresponding to the Portuguese noun `Tiago' produced by the same speakers, totalizing 2 items;
\item co-articulated vowels in a word context corresponding to the Portuguese verb `Saiu' produced by the same speakers, totalizing 2 items.
\end{itemize}
As indicated earlier, for each one of the eight original audio items, three modified versions were created by replacing, in the original audio items, the original voiced regions by synthetic voicing generated by each one of the discussed synthetic voicing techniques. Thus, in total, 32 audio files were organized in 8 listening tasks (four versions of two vowels and two words produced by a male speaker and a female speaker).

All audio files were equalized in loudness given that original recordings showed some differences and this could introduce some bias affecting subjective judgement. The Adobe Audition platform was used to set the loudness level of all audio files to -15 LUFS according to the ITU-R BS.1770-2 specification.

The subjective grading adopted in the listening tests follows the ITU-R BS.1116 Recommendation for subjective assessment of small impairments \cite{itu:BS1116} and the 100-point Continuous Quality Scale (CQS), as specified in Recommendation ITU-R BS.1284  \cite{itu:BS1284}. This scale adopts the following ranges in the grading of the sound differences heard between two audio items when one of them is taken as the original reference of highest quality: imperceptible (80\%-100\%), perceptible but not annoying (60\%-80\%), slightly annoying (40\%-60\%), annoying (20\%-40\%), and very annoying (0\%-20\%).

Listeners were advised to use a high-quality headphone in the different listening tasks given that some of the impairments are rather subtle. 

In each one of the 8 listening tasks, the sequence (A, B, C, D) of the four versions of each audio item was randomized, which means that the reference was hidden. Listeners were instructed to listen to all 4 versions and, based on a judgement combining auditory impression of quality and naturalness, listeners were asked to decide first what version corresponds to the original version (or reference). Then, listeners were asked to assign the maximum grade (100\%) to that version, and to grade the remaining three versions according to the differences heard (in terms of quality and naturalness, and using a number between 0\% and 99\%) between the chosen reference and each version. In each listening task, the grading scale was permanently visible so as to facilitate the grading decisions. The 8 listening tasks were organized in a PPT document and subjective gradings were inserted in a XLS containing the same organization and instructions  of the PPT document. This document containing all 32 audio files is available for download\footnote{{\tiny{\tt \url{https://github.com/Anibal-Ferreira/FreTimGlo_subjTEST/}}}}.

Thirty volunteer subjects were recruited to participate in the listening tests, 8 female and 22 male subjects. Volunteers were either university graduate students or professors in the areas of speech/audio research or Speech and Language Therapy. Their average age was 40.7, the standard deviation was 10.9, the maximum age was 65 years, and the minimum age was 23. Fourteen subjects indicated to possess a music background and none reported having hearing problems.

\begin{figure}[htb]
\centering
\includegraphics[width=0.85\columnwidth]{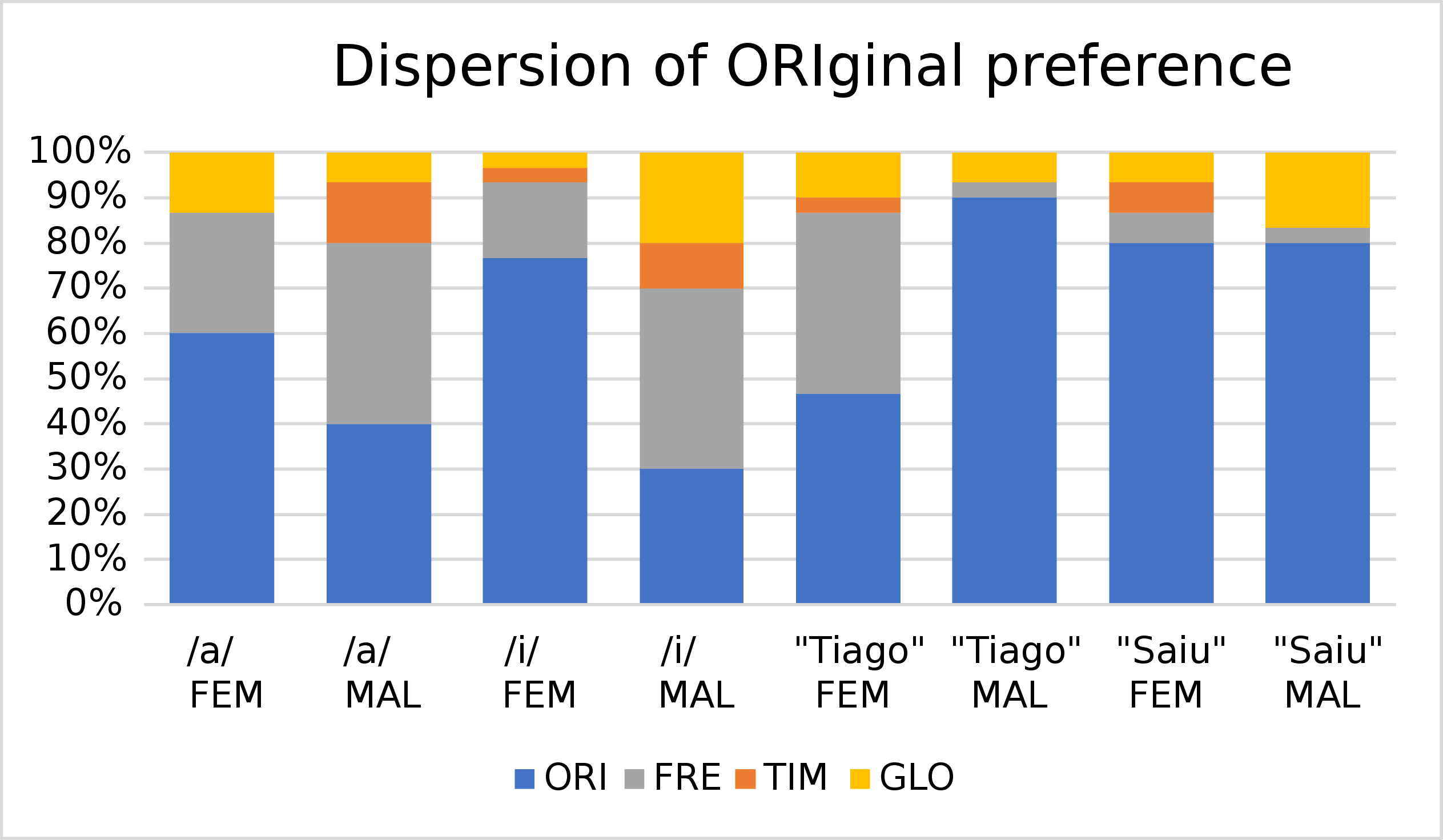}
\caption{For each one of the 8 listening tasks, the vertical bar represents the percent dispersion among all four audio versions (ORI, FRE, TIM, GLO) of the perceived reference audio file. True positives are represented by the blue bar segments.}
\label{fig:dispersion_3}
\end{figure}

We first present the results of the listening tests regarding the correct identification, in each listening task, of the reference audio file. As this reference was randomized with the other versions, i.e., it was hidden, it is of interest to asses the extent to which it is clearly discernible among the different versions. A correct identification percentage close to 100\% indicates that the processed versions exhibit clear audible differences, which represents evidence that the synthetic voicing is of poor quality. Conversely, a correct identification percentage significantly lower than 100\% indicates that the audio differences between reference and processed versions are not perceptually obvious, which represents evidence that the synthetic voicing has high quality and naturalness. Figure \ref{fig:dispersion_3} shows the dispersion, in each listening task, of the perceived reference among the four possible alternatives (ORI, TIM, FRE, and GLO).
In each listening task, the percentage of true positives is represented by a blue bar segment.

Regarding the listening tests involving sustained vowels, we observe that:
\begin{itemize}
\item first, as the true positives range from 30\% to 75\%, the three synthetic voicing alternatives show an interesting potential in delivering high quality and natural sounding vowels; 
\item second, regarding the male /a/ and /i/ sustained vowels, listeners confused the `FRE' with the `ORI' reference file (40\% vs. 40\% for the /a/ vowel, and 40\% vs. 30\% for the /i/ vowel), which indicates that frequency-domain synthesis delivers high quality confirming the perceptual assessment conclusions in \cite{speech:ajf2020_1, speech:silva2020};
\item finally, for all sustained vowels, the frequency-domain synthesis of voicing (`FRE'  version) is confused more with the reference ORI, and this is followed by the physiologically inspired time-domain synthesis of voicing (`GLO'  version), and finally by the combined frequency and time-domain synthesis of voicing (`TIM'  version); this suggests that the synthesis procedure detailed in Subsection \ref{sec:TDsyntOAPP} is capable of generating high-quality and natural sounding voicing of sustained vowels.
\end{itemize}
Concerning the listening tests involving co-articulated vowels in word context, we note that:
\begin{itemize}
\item except for one particular case (`Tiago'-FEM), true positives are in the order of 80\% or higher, which indicates that noticeable differences exist between the reference and processed versions;
\item the `GLO' version was confused with the `ORI' reference more often than any of the remaining two synthetic voicing alternatives, which indicates a clear human preference for the quality and naturalness delivered by the `GLO' synthetic voicing;
\item of all the synthetic voicing alternatives, the combined frequency and time-domain synthesis of voicing (`TIM'  version) was
selected less often than any other processing alternative.
\end{itemize}
Overall, the results shown in Fig. \ref{fig:dispersion_3} indicate that, in the case of sustained vowels, the sound quality and naturalness obtained by frequency-domain synthesis of voicing is preferred by human listeners, even surpassing the appreciation of the original audio references. In the case of co-articulated vowels in word context that does not hold true. Instead, except for the `Tiago'-FEM case, the time-domain synthesis of voicing (i.e., the `GLO'  version) appears to be more effective in capturing the preference of human listeners, although in this case listeners are more consistent in correctly identifying the original audio references. It is hypothesized that both of these outcomes are explained by the fact that the original audio files, i.e., the references, may not sound as `clean' as some of the processed versions. The detailed listening test results discussed next may shed some light on this.

\begin{figure}[htb]
\centering
\includegraphics[width=0.85\columnwidth]{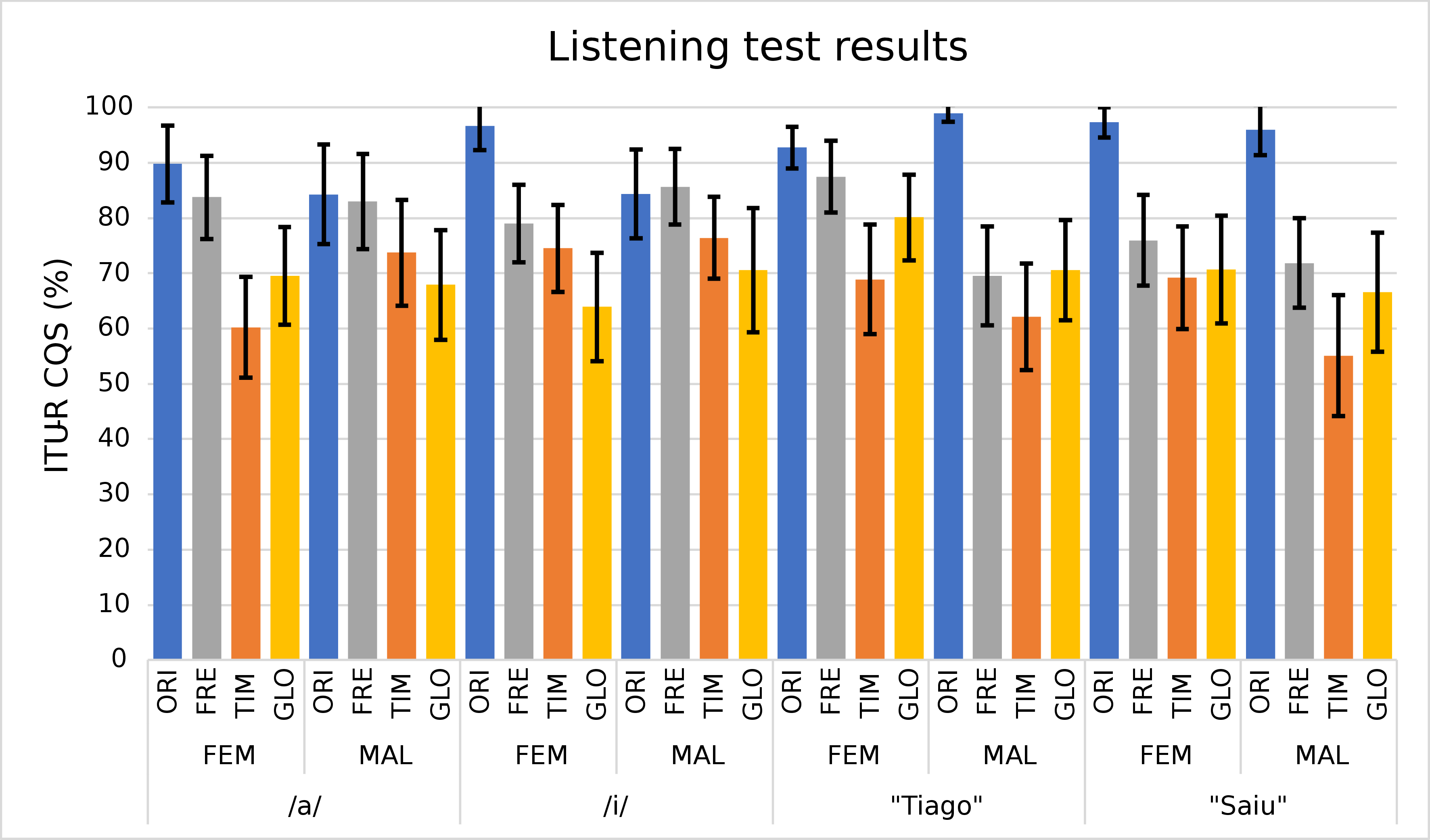}
\caption{Means and associated 95\% confidence intervals of the listening test scores on the quality and naturalness of synthetic voicing obtained with frequency-domain (FRE), combined frequency and time domain (TIM), and physiologically inspired time-domain (GLO) techniques.}
\label{fig:listening_4}
\end{figure}

The listening test results are shown in Fig. \ref{fig:listening_4}.
This figure represents the means and associated 95\% confidence intervals emerging from the scores due to each of the eight listening tasks, four of them involving sustained vowels, and four of them involving co-articulated vowels in word context.

In the following, we analyze pairs of tests per vowel (/a/ and /i/), and word (`Tiago' and `Saiu').

\subsubsection{Sustained /a/ vowel tests}
\label{sec:listening_a}

Tables \ref{tab:FEMa} and \ref{tab:MALa} detail the numerical results displayed graphically in Fig. \ref{fig:listening_4} concerning the female and male sustained /a/ vowel tests, respectively. 
\begin{table}[htb]
\centering
\begin{tabular}{c | c | c | c | c}
 {\color{NavyBlue}\footnotesize FEM /a/} & {\small ORI} & {\small FRE} & {\small TIM} & {\small GLO} \\ 
 \hline
{\small ORI} & 89.8 & {\color{BrickRed}\small{\it p}=0.31} & {\color{BrickRed}\small\bf{\it p}=3.5E-7}  & {\color{BrickRed}\small\bf{\it p}=0.002}  \\
{\small FRE} & - & 87.7 & {\color{BrickRed}\small\bf{\it p}=1.5E-5}  & {\color{BrickRed}\small\bf{\it p}=1.8E-3}\\
{\small TIM} & - & - & 60.2 & {\color{BrickRed}\small{\it p}=0.053} \\
{\small GLO} & - & - & - & 69.6  \\
\end{tabular}
\caption{Average ITU-R CQS scores (in \%) and {\it p}-values of a two-tailed {\it t}-test comparing scores pertaining to the FEM /a/ sustained vowel test. A bold font means statistically significant differences, {\it df}=29.}
\label{tab:FEMa}
\end{table}

\begin{table}[htb]
\centering
\begin{tabular}{c | c | c | c | c}
 {\color{NavyBlue}\footnotesize MAL /a/} & {\small ORI} & {\small FRE} & {\small TIM} & {\small GLO} \\ 
 \hline
{\small ORI} & 84.2 & {\color{BrickRed}\small{\it p}=0.80} & {\color{BrickRed}\small{\it p}=0.084}  & {\color{BrickRed}\small\bf{\it p}=0.018}  \\
{\small FRE} & - & 83.0 & {\color{BrickRed}\small{\it p}=0.09}  & {\color{BrickRed}\small\bf{\it p}=0.017}\\
{\small TIM} & - & - & 73.7 & {\color{BrickRed}\small{\it p}=0.33} \\
{\small GLO} & - & - & - & 67.9  \\
\end{tabular}
\caption{Average ITU-R CQS scores (in \%) and {\it p}-values of a two-tailed {\it t}-test comparing scores pertaining to the MAL /a/ sustained vowel test. A bold font means statistically significant differences, {\it df}=29.}
\label{tab:MALa}
\end{table}
In both cases, the average score associated with the original audio item (ORI) exceeds those associated with the three processed versions (FRE, TIM, GLO), as expected. However, a two-tailed {\it t}-test between pairs of means and considering a significance level of $\alpha = 0.05$ reveals that, relative to the reference (ORI), statistically significant differences are found with respect to the TIM and GLO versions in the case of the female audio items, and only with respect to the GLO version of the case of the male audio item.
This means that in both cases, frequency-domain processing delivers a quality that is not statistically different from that of the original audio items, which confirms the conclusion in \cite{speech:silva2020} using a similar frequency-domain processing and sustained spoken and sung vowels.

On the other hand, with respect to the TIM and GLO versions, listeners found that the differences heard are `perceptible but not annoying'. Comparing the scores of these versions with the FRE scores, in both cases statistically significant differences exist between the FREQ and GLO versions, but not between the TIM and GLO versions. It is interesting to note that the average score of the GLO version exceeds that of the TIM version in the case of the FEM audio item, and the opposite happens in the case of the MAL audio item. Thus, the sustained /a/ vowel listening test results indicate that signal alterations due to frequency-domain processing are imperceptible, and that signal alterations associated with the TIM or GLO versions are noticeable with statistical significance, although not annoying. However, results are not conclusive as to which TIM or GLO version is preferable perceptually.

\subsubsection{Sustained /i/ vowel tests}
\label{sec:listening_i}

Tables \ref{tab:FEMi} and \ref{tab:MALi} clarify the numerical results displayed graphically in Fig. \ref{fig:listening_4} regarding the female and male sustained /i/ vowel tests, respectively. 
\begin{table}[htb]
\centering
\begin{tabular}{c | c | c | c | c}
 {\color{NavyBlue}\footnotesize FEM /i/} & {\small ORI} & {\small FRE} & {\small TIM} & {\small GLO} \\ 
 \hline
{\small ORI} & 96.7 & {\color{BrickRed}\small\bf{\it p}=2.9E-4} & {\color{BrickRed}\small\bf{\it p}=5.1E-5}  & {\color{BrickRed}\small\bf{\it p}=3.3E-7}  \\
{\small FRE} & - & 79.0 & {\color{BrickRed}\small{\it p}=0.22}  & {\color{BrickRed}\small\bf{\it p}=4.5E-4}\\
{\small TIM} & - & - & 74.5 & {\color{BrickRed}\small\bf{\it p}=0.012} \\
{\small GLO} & - & - & - & 63.9  \\
\end{tabular}
\caption{Average ITU-R CQS scores (in \%) and {\it p}-values of a two-tailed {\it t}-test comparing scores pertaining to the FEM /i/ sustained vowel test. A bold font means statistically significant differences, {\it df}=29.}
\label{tab:FEMi}
\end{table}

\begin{table}[htb]
\centering
\begin{tabular}{c | c | c | c | c}
 {\color{NavyBlue}\footnotesize MAL /i/} & {\small ORI} & {\small FRE} & {\small TIM} & {\small GLO} \\ 
 \hline
{\small ORI} & 84.4 & {\color{BrickRed}\small{\it p}=0.81} & {\color{BrickRed}\small{\it p}=0.13}  & {\color{BrickRed}\small{\it p}=0.052}  \\
{\small FRE} & - & 85.7 & {\color{BrickRed}\small\bf{\it p}=0.045}  & {\color{BrickRed}\small\bf{\it p}=0.0079}\\
{\small TIM} & - & - & 76.4 & {\color{BrickRed}\small{\it p}=0.31} \\
{\small GLO} & - & - & - & 70.6  \\
\end{tabular}
\caption{Average ITU-R CQS scores (in \%) and {\it p}-values of a two-tailed {\it t}-test comparing scores pertaining to the MAL /i/ sustained vowel test. A bold font means statistically significant differences, {\it df}=29.}
\label{tab:MALi}
\end{table}
A clear outcome stands out: relative to the female reference audio item (ORI), the three voicing versions are perceived as showing audible differences, with statistical significance, although not annoying. However, that does not hold with respect to the male reference audio item. This can be looked at as a consequence of the fact that, as noted earlier, and as shown in Fig. \ref{fig:dispersion_3}, the true positives in the identification of the reference /i/ FEM vowel are significantly higher than in the case of the reference /i/ MAL vowel. This also means that the three types of synthetic voicing sound much more natural in the case of the MAL /i/ listening task than in the case of the FEM /i/ listening task. In addition, with statistical significance, the GLO version is perceived as being noticeably worse than the remaining versions in the case of female listening task, but not in the case of the male listening task.

Thus, the sustained /i/ vowel listening test results indicate that audible differences between the reference and the various voicing alternatives are not always clear with statistical significance, but when they exist, they are regarded as mild and not annoying.

\subsubsection{`Tiago' word tests}
\label{sec:Tiago_word}

Tables \ref{tab:FEMtiago} and \ref{tab:MALtiago} show the numerical results displayed graphically in Fig. \ref{fig:listening_4} regarding the female and male co-articulated /i--a/ vowels tests in the `Tiago' word context, respectively. 
\begin{table}[htb]
\centering
\begin{tabular}{c | c | c | c | c}
 {\color{NavyBlue}\footnotesize FEM `Tiago'} & {\small ORI} & {\small FRE} & {\small TIM} & {\small GLO} \\ 
 \hline
{\small ORI} & 92.7 & {\color{BrickRed}\small{\it p}=0.091} & {\color{BrickRed}\small\bf{\it p}=3.0E-5}  & {\color{BrickRed}\small\bf{\it p}=0.0082}  \\
{\small FRE} & - & 87.5 & {\color{BrickRed}\small\bf{\it p}=3.7E-4}  & {\color{BrickRed}\small{\it p}=0.15}\\
{\small TIM} & - & - & 68.9 & {\color{BrickRed}\small\bf{\it p}=0.0011} \\
{\small GLO} & - & - & - & 80.1  \\
\end{tabular}
\caption{Average ITU-R CQS scores (in \%) and {\it p}-values of a two-tailed {\it t}-test comparing scores pertaining to the FEM `Tiago' word test. A bold font means statistically significant differences, {\it df}=29.}
\label{tab:FEMtiago}
\end{table}

\begin{table}[htb]
\centering
\begin{tabular}{c | c | c | c | c}
 {\color{NavyBlue}\footnotesize MAL `Tiago'} & {\small ORI} & {\small FRE} & {\small TIM} & {\small GLO} \\ 
 \hline
{\small ORI} & 99.0 & {\color{BrickRed}\small\bf{\it p}=1.4E-9} & {\color{BrickRed}\small\bf{\it p}=1.6E-8}  & {\color{BrickRed}\small\bf{\it p}=1.1E-6}  \\
{\small FRE} & - & 69.6 & {\color{BrickRed}\small{\it p}=0.077}  & {\color{BrickRed}\small{\it p}=0.83}\\
{\small TIM} & - & - & 62.2 & {\color{BrickRed}\small\bf{\it p}=0.014} \\
{\small GLO} & - & - & - & 70.6  \\
\end{tabular}
\caption{Average ITU-R CQS scores (in \%) and {\it p}-values of a two-tailed {\it t}-test comparing scores pertaining to the MAL `Tiago' word test. A bold font means statistically significant differences, {\it df}=29.}
\label{tab:MALtiago}
\end{table}
The FEM `Tiago' scores present a similar relative ranking when compared to those of the FEM /a/ scores, but a significant increase is observed in the scores of the TIM version and, especially, in the scores of the GLO version (80.1\% vs. 69.6\%), which suggest that the physiological-inspired GLO version is more adequate to handle co-articulation than the TIM version. This is confirmed by the fact that the average score difference between the ORI and the GLO versions is statistically significant (ORI better than GLO, as expected), as well as the average score difference between the GLO and TIM versions (GLO better than TIM), but the average score difference between the FRE and GLO version is not statistically significant.

In the case of the MAL `Tiago' test, the GLO average score surpasses those of the TIM and FRE versions, although not with statistical significance in the latter case. This outcome is particularly revealing because the MAL `Tiago' test is the case where listeners more reliably identified the reference, as noted earlier and shown in Fig. \ref{fig:dispersion_3}.

Thus, the `Tiago' test results indicate that the GLO scores rank better than the TIM scores with statistical significance, but score differences between the GLO and FRE versions are not statistically significant, although the perceived audio differences with respect to the reference are categorized as mild and not annoying.

\subsubsection{`Saiu' word tests}
\label{sec:Saiu_word}

Tables \ref{tab:FEMsaiu} and \ref{tab:MALsaiu} show the numerical results displayed graphically in Fig. \ref{fig:listening_4} regarding the female and male co-articulated /a--i--u/ vowels tests in the `Saiu' word context, respectively. 
\begin{table}[htb]
\centering
\begin{tabular}{c | c | c | c | c}
 {\color{NavyBlue}\footnotesize FEM `Saiu'} & {\small ORI} & {\small FRE} & {\small TIM} & {\small GLO} \\ 
 \hline
{\small ORI} & 97.3 & {\color{BrickRed}\small\bf{\it p}=5.9E-6} & {\color{BrickRed}\small\bf{\it p}=2.5E-6}  & {\color{BrickRed}\small\bf{\it p}=1.4E-5}  \\
{\small FRE} & - & 75.9 & {\color{BrickRed}\small{\it p}=0.062}  & {\color{BrickRed}\small{\it p}=0.24}\\
{\small TIM} & - & - & 69.2 & {\color{BrickRed}\small{\it p}=0.74} \\
{\small GLO} & - & - & - & 70.7  \\
\end{tabular}
\caption{Average ITU-R CQS scores (in \%) and {\it p}-values of a two-tailed {\it t}-test comparing scores pertaining to the FEM `Saiu' word test. A bold font means statistically significant differences, {\it df}=29.}
\label{tab:FEMsaiu}
\end{table}

\begin{table}[htb]
\centering
\begin{tabular}{c | c | c | c | c}
 {\color{NavyBlue}\footnotesize MAL `Saiu'} & {\small ORI} & {\small FRE} & {\small TIM} & {\small GLO} \\ 
 \hline
{\small ORI} & 96.0 & {\color{BrickRed}\small\bf{\it p}=4.1E-5} & {\color{BrickRed}\small\bf{\it p}=1.8E-7}  & {\color{BrickRed}\small\bf{\it p}=2.1E-5}  \\
{\small FRE} & - & 71.9 & {\color{BrickRed}\small\bf{\it p}=1.0E-4}  & {\color{BrickRed}\small{\it p}=0.16}\\
{\small TIM} & - & - & 55.1 & {\color{BrickRed}\small\bf{\it p}=7.0E-4} \\
{\small GLO} & - & - & - & 66.6  \\
\end{tabular}
\caption{Average ITU-R CQS scores (in \%) and {\it p}-values of a two-tailed {\it t}-test comparing scores pertaining to the MAL `Saiu' word test. A bold font means statistically significant differences, {\it df}=29.}
\label{tab:MALsaiu}
\end{table}
It is clear that the FEM `Saiu' tests scores follow the following (not statistically significant) decreasing order: FRE $>$ GLO $>$ TIM. In all three cases, audible differences with respect to the ORI version are classified by the listeners as perceptible but not annoying. A clear conclusion from the MAL `Saiu' tests is that this is the case where the average score of one of the versions (TIM) is the lowest of all tests (55.1\%) and it differs with statistical significance from the remaining versions (ORI, FRE, GLO).
A relevant conclusion from both FEM and MAL `Saiu' tests is that although the average scores associated with the FRE version exceed those associated with the GLO version, differences are not statistically significant.

In the following, we summarize the main conclusions emerging from the listening tests.
\begin{itemize}
\item In the case of co-articulated vowels, listeners are able to clearly identify the hidden reference in most cases, whereas in the case of sustained vowels the reference identification is much more difficult, and even poor in certain cases. This leads to the conclusion that synthetic voicing of sustained vowels is not as informative and realistic as synthetic voicing of co-articulated vowels. In fact, the latter scenario is much more representative of the challenging scenario of interest: synthetic voicing of running whispered speech. This is especially true when the general quality of the synthetic voicing discussed in this paper can be regarded as high given that in almost all test cases the audible differences between the original audio items and the synthetic versions are classified as perceptible but not annoying.
\item While in the case of sustained vowels the average scores of the combined frequency and time-domain voicing (TIM) tend to exceed those of the physiologically-inspired time-domain voicing (GLO) even though differences are not statistically significant in most cases, in the case co-articulated vowels the opposite takes places, i.e., the average scores of the physiologically-inspired time-domain voicing (GLO) tend to exceed those of the the combined frequency and time-domain voicing (TIM) and differences are statistically significant in most cases. This points to the conclusion that physiologically-inspired time-domain voicing (GLO) is more suited than the TIM alternative to implement  synthetic voicing in running whispered speech as this scenario involves mostly co-articulated vowels. 
\item In the case of sustained vowels, the average scores of the frequency-domain voicing (FRE) are higher, with statistical significance, than those of the physiologically-inspired time-domain voicing (GLO), whereas in the case of co-articulated vowels, differences are not statistically significant. In all cases, however, audible differences between the reference (ORI) and the FRE or GLO versions are categorized by the listeners as perceptible but not annoying. This not only corroborates the previous observations, but also suggests that the FRE and GLO voicing alternatives are better suited to implement synthetic voicing of whispered speech given that co-articulated vowels are much more likely to occur than sustained vowels.
\end{itemize}

To the best knowledge of the authors, only the work in \cite{speech:ajf2020_1} has implemented a similar frequency-domain synthetic voicing of sustained vowels (FRE, column ``C'' in Fig. 6 of \cite{speech:ajf2020_1}), and a similar combined frequency and time-domain synthetic voicing of sustained vowels (TIM, column ``F'' in Fig. 6 of \cite{speech:ajf2020_1}). The FRE results presented in \cite{speech:ajf2020_1} for spoken and sung sustained vowels are in the range 70\%--90\%, quite in line with the results in this paper for sustained (spoken) vowels. On the other hand, the listening test results presented in \cite{speech:ajf2020_1} regarding the TIM synthetic voicing alternative are in the range 55\%--70\% for spoken vowels, and 45\%--55\% for sung vowels, whereas the results in this paper for (spoken) vowels are the range 60\%--77\%. Is it believed that the observed improvements can be explained by the fact that in this paper the TIM synthetic voicing includes an additional post-processing step smoothing the transitions between adjacent pitch pulses, as explained in Subsection \ref{sec:TDsyntConPP}.

\section{Conclusions and future work}
\label{sec:conclusion}

In this paper, we addressed two topics that are intertwined and that play a key role in preparing the terrain to the restoration of natural speech from whispered speech: pitch pulse segmentation in voiced regions of sustained and co-articulated vowels, and comparison of three synthetic voicing techniques, one of them being inspired by the physiological mechanisms of voice production.

The literature was reviewed regarding the first topic and it was concluded that in order to study in detail the time and spectral evolution of the phase/magnitude structure of adjacent pitch pulses, a new pitch pulse segmentation approach is required that focuses on pitch pulse periodicity rather than on epochs. A new technique that relies on the analysis of the phase of the fundamental frequency was presented and experimental results were shown. These results revealed that the harmonic phase/magnitude structures of the same vowels produced by the same speakers may differ depending on the sustained and co-articulation production regime. Still, provided that representative phase/magnitude vowel templates are available, entire co-articulation regions may be synthesized as a result of pitch pulse-based interpolation between those templates.

A literature review regarding the topic on the conversion of whispered speech to natural speech highlighted two main approaches: model-based and data-driven. The main advantages of model-based approaches are: they are interpretable because they rely strongly on the established source-filter voice production paradigm, they are modest in computational requirements, which not only facilitates lightweight implementation on portable devices, but also facilitates compliance with the real-time and on-the-fly operational requirements. However, in most cases, model-based approaches are critically dependent on the architecture of specific codecs, or rigid filter models, that strongly limit the independent control of the glottal excitation and vocal tract filter contributions, with critical impacts in terms of control of prosody, naturalness, idiosyncratic sound signature, and overall signal quality.

On the other hand, data-driven approaches may deliver good voice conversion results but, typically, they require massive amounts of paired whispered speech and natural speech data, a sophisticated vocoder for signal reconstruction, and heavy computational resources that quite likely prohibit lightweight implementation on portable devices, in addition to compliance with real-time and on-the-fly operational requirements.
 
Given our focus on lightweight algorithms capable of operating in real-time and on-the-fly, we proposed and compared three synthetic voicing algorithms that leverage previous results in this area. All algorithms use parametric models to govern the synthesis process. One algorithm operates in the frequency-domain (FRE) and uses pitch and harmonic magnitude and phase models, a second algorithm uses the same parametric information to synthesize, and concatenate individual pitch pulses in the time domain (TIM), and a third algorithm uses only pitch and harmonic magnitude information to drive a physiologically inspired synthetic voicing process (GLO) by filtering individual glottal excitation pulses and overlapping and adding the individual filtered outputs.

The three algorithms were assessed in an objective perspective by comparing the spectrograms of synthetic voicing of sustained and co-articulated vowels. Is general, it was concluded that the original spectral structure of the original signals was well preserved in the processed signals irrespective of the synthesis method, and that time-domain synthesis (TIM and GLO algorithm) give rise to spectrogram showing a cleaner harmonic continuity, in particular of higher-order partials.

The listening tests results were presented and discussed. As the reference was randomized with the other stimuli in these tests, it was possible to conclude that listeners identify correctly the reference much more easily in the case of the co-articulated vowels listening tasks than in the case of the sustained vowels listening tasks. This supports the conclusion that co-articulated vowels are more informative in characterizing the perceptual impact of different synthetic voicing techniques.

Overall, the listening tests indicate that the FRE and GLO algorithms are the most promising synthetic voicing techniques as the co-articulated vowels scores characterizing signal quality and naturalness are not different with statistical significance, and audible differences with respect to the reference audio files were categorized as perceptible but not annoying. In particular, the `GLO' synthetic voicing technique has shown great potential, especially with respect to male voice signals.  

Future work towards a complete lightweight algorithm implementing restoration of natural speech from whispered speech will tackle two main challenges. A first challenge  involves leveraging the results and insights emerging from this paper, with respect to the physiologically inspired synthetic voicing technique (GLO), notably by using a glottal pulse model that is grounded on true physiological data, rather than on an idealized mathematical model such as the Liljencrants-Fant model. A parametric model of such physiologically-inspired glottal pulse will be sought that allows the synthesis of individual glottal pulses in a flexible way accommodating the possibility of adjusting the waveform shape of each individual pulse to a desired phonation type (e.g., tense, modal, or breathy), and including natural glottal phenomena such as extra glottal pulses in a single glottal period. By explicitly decoupling the influences of the glottal excitation and vocal tract filtering in synthetic voicing, additional degrees of freedom are available to tailor idiosyncratic voice attributes that characterize a desired individual sound signature.

A second challenge has to do with the accurate phonetic-oriented segmentation of whispered speech, and prosody estimation. Phonetic-oriented segmentation is key in implanting synthetic voicing in those phonemic regions in whispered speech that are voiced in natural speech. Significant research work has been carried out recently in this area \cite{voice:goncalo2023, voice:costa2021}.

Following phonetic-oriented segmentation, prosody estimation and control determines voice sound individuality and overall speech expressivity and naturalness. While any one of the discussed voicing algorithms is capable of synthesizing a desired $f_{0}$ contour, estimating that contour, i.e., the prosody, from whispered speech, remains a challenge. Earlier works in this area \cite{voice:fujisaki1984, speech:ian2015} will be reviewed and the most promising techniques will be adapted and perfected.

Both phonetic-oriented segmentation and prosody estimation techniques will be implemented fulfilling the three essential requirements of the desired whispered speech to natural/voiced speech restoration algorithm: i) convincing synthetic speech quality and naturalness, ii) lightweight computational complexity, and iii) operation in real-time and on-the-fly.

\bibliographystyle{elsarticle-num} 



\bibliography{../../../livros/BOOK/bib/mpeg,../../../livros/BOOK/bib/audiocompr,../../../livros/BOOK/bib/psycho,../../../livros/BOOK/bib/signalproc,../../../livros/BOOK/bib/filterdesign,../../../livros/BOOK/bib/pitch,../../../livros/BOOK/bib/speech,../../../livros/BOOK/bib/voice,../../../livros/BOOK/bib/forensics,../../../livros/BOOK/bib/itu}

\end{document}